\documentclass[aps,pra,reprint,superscriptaddress,longbibliography,twocolumn]{revtex4-2}
\usepackage[T1]{fontenc}
\usepackage[utf8]{inputenc}
\usepackage{lmodern}
\usepackage{microtype}
\usepackage{mathtools}
\usepackage{graphicx}
\usepackage{dcolumn}
\usepackage{physics}
\usepackage{amsmath,amssymb,amsthm,amsfonts}
\usepackage{bm,bbm}
\usepackage{lipsum,babel}
\usepackage[normalem]{ulem}
\usepackage{tabularx}
\usepackage{algorithm}            
\usepackage{algpseudocode}        
\usepackage{amsmath}        
\usepackage{xcolor}
\usepackage{overpic}
\usepackage{hyperref} 
\usepackage[table]{xcolor}
\DeclareMathOperator*{\argmax}{arg\,max}

\newcommand{\actNone}{00}
\newcommand{\actZ}{10}
\newcommand{\actX}{01}
\newcommand{\actB}{11}
\newcommand{\ZZ}{\mathbb Z}
\renewcommand{\c}{\sigma^2}
\newcommand{\h}{\sigma_{\mathrm{idle}}^2}
\newcommand{\g}{\sigma_{\mathrm{gkp}}^2}

\allowdisplaybreaks[1]

\begin{document}

\title{When to Skip Syndrome Extraction in Surface-GKP Codes}

\author{Vaughn Sohn}
\affiliation{Department of Physics, Korea Advanced Institute of Science and Technology, Daejeon 34141, Korea}
\affiliation{Department of Electrical Engineering, Korea Advanced Institute of Science and Technology, Daejeon 34141, Korea}

\author{Changhun Oh}
\email{changhun0218@gmail.com}
\affiliation{Department of Physics, Korea Advanced Institute of Science and Technology, Daejeon 34141, Korea}

\begin{abstract}
    Fault-tolerant quantum error correction requires repeated syndrome extraction to address errors induced by the syndrome-extraction circuit itself. 
    However, repeated syndrome extraction incurs significant overhead in terms of gate count and ancilla consumption (e.g., Gottesman-Kitaev-Preskill~(GKP)~states). 
    Moreover, noisy syndrome extraction can itself inject additional errors into the data qubits. 
    To address these issues, we propose a concrete adaptive skipping scheme for the surface-GKP code, a representative GKP-concatenated architecture, that uses analog information naturally generated during inner GKP correction. 
    At each round, the scheme selects one of four actions: measuring both $Z$-type and $X$-type surface-code stabilizers, measuring only one type, or skipping both types and reusing previous syndromes. 
    The decision is based on a reliability comparison between reusing the previous syndrome value and performing a new noisy syndrome extraction. 
    Using circuit-level simulations, we show that the adaptive skipping scheme can reduce the number of surface-code stabilizer measurements while maintaining logical error rates comparable to or lower than those of the full-measurement baseline. 
    The improvement is most pronounced when gate and measurement noise are larger than idle noise, so that avoiding unnecessary syndrome extraction reduces the noise injected into the code. 
    These results indicate that analog information from inner GKP correction can be used not only to improve decoding but also to reduce the measurement overhead of outer-code syndrome extraction.    
\end{abstract}
\maketitle

\section{Introduction}\label{sec:introduction} 
Fault-tolerant quantum computers are expected to provide computational advantages for tasks that are challenging for classical computers, including integer factorization, unstructured search, and the simulation of quantum systems~\cite{shor_Algorithms_1994, shor_PolynomialTime_1997, grover_fast_1996, feynman_Simulating_1982, lloyd_Universal_1996}. 
Realizing these advantages on noisy hardware requires quantum error correction~(QEC) to suppress logical error rates. 
Recent theoretical advances and experimental demonstrations have therefore accelerated efforts to implement QEC across a wide range of quantum-computing platforms~\cite{putterman_Hardwareefficient_2025, reichardt_Faulttolerant_2025, acharya_Quantum_2025, bluvstein_Logical_2024, paetznick_Demonstration_2024, matsos_Robust_2024, reichardt_Demonstration_2024, sundaresan_Demonstrating_2023, sivak_Realtime_2023, krinner_Realizing_2022}.

A central obstacle to fault-tolerant QEC is that syndrome extraction is itself noisy. 
Each extraction round involves ancilla preparation, gate operations, and measurements, all of which can introduce faults. 
Consequently, the syndrome obtained from a single round can be ambiguous: a nontrivial outcome may reflect either an error on the data qubits or a fault in the extraction circuit. 
Standard fault-tolerant protocols resolve this ambiguity by repeating syndrome extraction over multiple rounds. 
This repetition, however, increases both the QEC cycle time and the resources required for syndrome extraction, including gate operations and ancilla-state preparations. 
Moreover, because each additional extraction round is itself noisy, repeated extraction can also inject further errors into the data. 
This motivates approaches that reduce the cost of obtaining reliable syndrome information.

Several approaches have been proposed to reduce the cost of repeated syndrome extraction, including single-shot QEC, flag-qubit protocols, and autonomous QEC~\cite{quintavalle_SingleShot_2021, bombin_SingleShot_2015, chao_Flag_2020, chao_Quantum_2018, quirion_Autonomous_2024a, zeng_Approximate_2023}. 
In this work, we focus on a complementary direction: adaptive syndrome extraction. 
Adaptive strategies use information obtained during error correction to decide how many, or which, stabilizer measurements should be performed~\cite{berthusen_Adaptive_2025, anker_Compressing_2025, liou_Reducing_2025, tansuwannont_Adaptive_2023a, bhambay_Adaptive_2026, gupta2026boundaryawarestabilizerschedulingdistributed}. 
A representative example is Ref.~\cite{berthusen_Adaptive_2025}, which considered concatenated codes and used the syndrome outcomes of a low-overhead inner error-detecting code to determine which outer-code stabilizer generators should be measured. 
The underlying principle is that low-overhead information acquired during error correction can be used to avoid syndrome measurements that are unlikely to improve decoding.

This principle is particularly well suited to continuous-variable--discrete-variable~(CV--DV) concatenated architectures, where measurement outcomes are continuous before they are converted into discrete error or syndrome estimates. 
These continuous outcomes provide reliability information for the corresponding discrete decisions: an outcome well within a decision region supports a reliable estimate, whereas an outcome close to a boundary between two possible decisions indicates greater ambiguity. 
Such analog, or soft, information has already been shown to improve GKP and GKP-concatenated decoding by entering decoder weights~\cite{fukui_Analog_2017, fukui_HighThreshold_2018, noh_Faulttolerant_2020a, berent_Analog_2024}. 
Here, we use the same information for a complementary purpose: deciding which outer-code syndrome measurements should be performed.
In GKP-concatenated architectures, this information is intrinsic rather than supplied by an auxiliary code: GKP qubits form the inner code, and routine GKP stabilizer measurements produce analog outcomes during inner-code correction.

We instantiate this principle in the surface-GKP code, a representative and practically relevant testbed for adaptive syndrome extraction in GKP-concatenated architectures~\cite{noh_Faulttolerant_2020a}. 
In this architecture, GKP qubits serve as the inner code, and the surface code serves as the outer code. 
Because GKP correction is imperfect in the presence of finite squeezing and circuit-level noise, residual GKP-level Pauli errors can remain after inner-code correction. 
These errors act as physical Pauli errors for the outer surface code and are detected through repeated surface-code syndrome extraction and corrected by outer-code decoding.
However, each surface-code stabilizer measurement consumes GKP ancilla states and involves active operations that can inject additional noise into the data. 
Thus, surface-GKP codes provide a natural setting in which analog information from the inner code can make outer-code syndrome extraction adaptive, with the goal of reducing unnecessary stabilizer measurements without introducing an additional auxiliary code.

Based on this observation, we propose a concrete adaptive syndrome-skipping strategy for the surface-GKP code.
At each round, after inner GKP correction, the protocol selects one of four global actions: measuring only $Z$-type surface-code stabilizers, measuring only $X$-type surface-code stabilizers, measuring both types, or skipping both types. 
For each stabilizer type, the decision compares two reliability estimates: the probability that the previous syndrome remains valid, inferred from analog GKP outcomes on the data qubits, and the probability that a new noisy surface-code syndrome extraction would produce a correct syndrome outcome. 
If reusing the previous syndrome is estimated to be more reliable, the corresponding syndrome extraction is skipped and the previous value is reused.

Circuit-level simulations show that this concrete adaptive rule can substantially reduce the number of surface-code stabilizer measurements while maintaining logical error rates comparable to, and in some regimes lower than, the full-measurement baseline. 
The improvement is most pronounced when active syndrome-extraction noise is larger than idle noise, so that avoiding unnecessary extraction reduces both GKP-ancilla consumption and circuit-induced noise. 
Together, these results show that analog information generated during routine GKP correction can be used not only for decoding, but also for reducing the measurement overhead of outer-code syndrome extraction.

The remainder of this paper is organized as follows. 
Sec.~\ref{sec:background} briefly reviews the surface-GKP code and the concept of analog information. 
Sec.~\ref{sec:strategy} describes the proposed adaptive syndrome-skipping strategy. 
Sec.~\ref{sec:results} presents simulation results comparing the proposed method with standard repeated syndrome extraction. 
Additional simulation details and supplementary numerical results are provided in Appendices~\ref{app:detail} and \ref{app:result}. 
Finally, Sec.~\ref{sec:discussion_and_conclusion} summarizes the results and discusses future research directions.

\section{Background}\label{sec:background} 
This section introduces the background needed to understand the proposed adaptive syndrome-skipping strategy. 
We first review GKP codes and surface-GKP concatenated codes within the stabilizer formalism, establishing the notation used throughout this work. 
We then introduce analog information in CV systems, focusing on how it provides reliability estimates for GKP stabilizer measurement outcomes. 
Further details on GKP codes, Gaussian shift noise models, and concatenated GKP architectures can be found in Refs.~\cite{grimsmo2021gkp, hillmann2025bosonic}.

\subsection{GKP code}
\begin{figure*}[tb]
    \includegraphics[width=\textwidth]{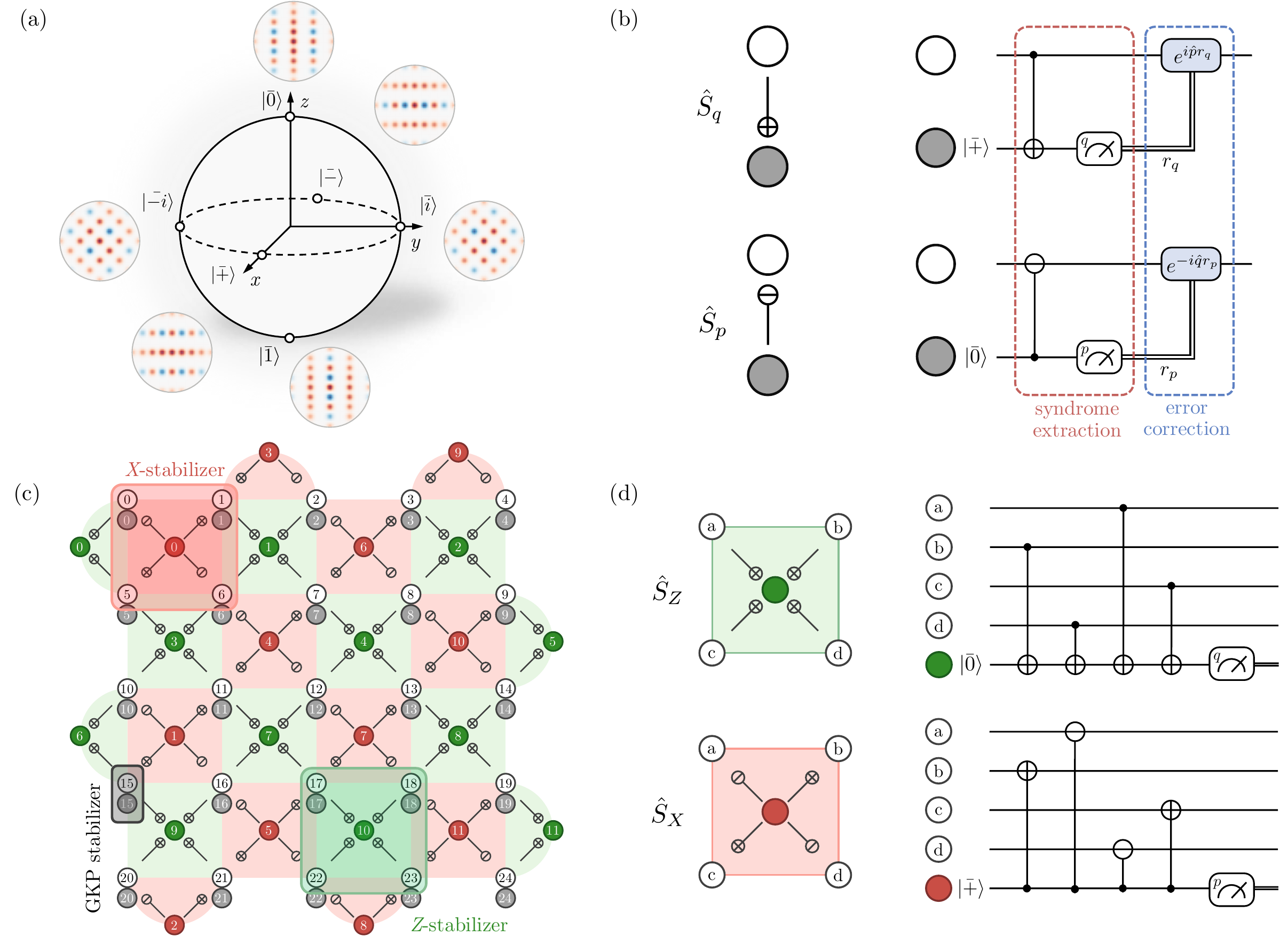}
    \caption{Surface-GKP code architecture. (a) Mapping of GKP states onto the Bloch sphere. Each logical state is illustrated as a Wigner function in phase space. (b) Circuit descriptions for GKP stabilizer measurement. The measurement symbols $q$ and $p$ denote homodyne measurement of the position and momentum quadratures, respectively. The symbol $\oplus$ denotes the SUM gate, $\ominus$ denotes the inverse-SUM gate, and the filled circle indicates the control mode. (c) Distance-5 surface-GKP code. White nodes denote data qubits. Grey, red, and green nodes represent ancilla qubits for GKP, $X$-type, and $Z$-type stabilizer measurements, respectively. (d) Circuit descriptions for surface-GKP stabilizer measurement.}
    \label{fig:surface-gkp}
\end{figure*}

The GKP code encodes a logical qubit into the infinite-dimensional Hilbert space of a quantum harmonic oscillator. 
Specifically, the square-lattice GKP code is defined as the simultaneous $+1$ eigenspace of the stabilizer group $\mathcal{S}_{\mathrm{gkp}}$, generated by the displacement operators: 
\begin{equation}
    \mathcal{S}_{\mathrm{gkp}} = \langle \hat{S}_q, \hat{S}_p \rangle, \quad \hat{S}_q \coloneqq e^{i 2 \sqrt{\pi} \hat{q}}, \quad \hat{S}_p \coloneqq e^{-i 2 \sqrt{\pi} \hat{p}}, 
\end{equation} 
where $\hat{q}$ and $\hat{p}$ denote the position and momentum operators, respectively. 
The logical Pauli operators are defined as 
\begin{equation} 
    \bar{Z}_{\mathrm{gkp}} \coloneqq e^{i \sqrt{\pi} \hat{q}}, \quad \bar{X}_{\mathrm{gkp}} \coloneqq e^{-i \sqrt{\pi} \hat{p}}. 
\end{equation} 
The logical basis states corresponding to the eigenstates of these operators are expressed as: 
\begin{equation} 
    \begin{aligned} 
        \ket{\bar{0}} &\coloneqq \sum_{n \in \mathbb{Z}} \ket{\hat{q} = 2 n \sqrt{\pi}}, \ \ket{\bar{1}} \coloneqq \sum_{n \in \mathbb{Z}} \ket{\hat{q} = (2 n + 1) \sqrt{\pi}}, \\ 
        \ket{\bar{+}} &\coloneqq \sum_{ n \in \mathbb{Z}} \ket{\hat{p} = 2 n \sqrt{\pi}}, \ \ket{\bar{-}} \coloneqq \sum_{n \in \mathbb{Z}} \ket{\hat{p} = (2 n + 1) \sqrt{\pi}}. 
    \end{aligned} 
\end{equation} 
As illustrated in Fig.~\ref{fig:surface-gkp}, the logical basis states form periodic combs in either position or momentum space, with lattice spacing $2\sqrt{\pi}$.

Ideal GKP states are superpositions of infinitely squeezed states, which are unphysical because they require infinite energy. 
For practical purposes, finite-energy GKP states are commonly modeled by applying a non-unitary Gaussian envelope operator to an ideal GKP state~\cite{matsuura_Equivalence_2020}:
\begin{equation}
    \ket{{\bar{\psi}}_\Delta} \propto e^{-\Delta \hat{n}} \ket{\bar{\psi}},
\end{equation}
where $\hat{n} = \hat{a}^\dagger \hat{a}$ denotes the number operator and $\Delta>0$ is the finite-energy parameter. 
The ideal GKP state is recovered in the limit $\Delta \to 0$; thus, the high-squeezing regime corresponds to small $\Delta$. 
In the position representation, the GKP wavefunction can be viewed as a comb of Gaussian peaks with finite width under a broad Gaussian envelope. 

To model shift errors in the GKP code, we use a Gaussian random displacement channel: 
\begin{equation} 
    \mathcal{N}[\sigma](\rho) \coloneqq \frac{1}{\pi \sigma^2} \int d^2 \alpha \exp\left({-\frac{|\alpha|^2}{\sigma^2}}\right) \hat{D}(\alpha) \rho \hat{D}^{\dagger}(\alpha), 
\end{equation} 
where $\hat{D}(\alpha) \coloneqq \exp(\alpha \hat{a}^\dagger - \alpha^* \hat{a})$ is the displacement operator. 
Physically, this channel introduces a random displacement $\alpha$ distributed according to a Gaussian distribution in phase space. 
Equivalently, writing $\alpha=(\xi_q+i\xi_p)/\sqrt{2}$, this operation corresponds to adding independent Gaussian noise $\xi_q, \xi_p \sim \mathcal{N}(0, \sigma^2)$ to the $\hat{q}$ and $\hat{p}$ quadratures, respectively. 

For sufficiently small $\Delta$, finite-squeezing effects can be approximated by a Gaussian random displacement channel with quadrature variance $\Delta/2$~\cite{matsuura_Equivalence_2020}.
This approximation replaces coherent finite-squeezing effects by an incoherent mixture of random displacement errors, providing a tractable effective noise model for approximate GKP states.
Consequently, the total effective displacement noise consists of intrinsic finite-squeezing noise and extrinsic channel noise.

Stabilizer measurements of $\hat{S}_q$ and $\hat{S}_p$ amount to measuring the quadratures $\hat q$ and $\hat p$ modulo $\sqrt{\pi}$.
Given a modular measurement outcome, we use nearest-lattice decoding, which selects the smallest displacement consistent with the syndrome.
This strategy is natural under Gaussian displacement noise, for which larger displacements are exponentially suppressed.
For a single quadrature, a displacement $\xi$ is decoded correctly whenever
\begin{equation}
    |\xi| < \frac{\sqrt{\pi}}{2}.
\end{equation}
Shifts outside this correctable region may be decoded as an equivalent displacement differing by an integer multiple of $\sqrt{\pi}$.
An odd multiple corresponds to a logical Pauli error.

In the circuit model, GKP stabilizers are measured using an ancilla GKP state, a SUM gate, and homodyne detection, as shown in Fig.~\ref{fig:surface-gkp}.
The SUM gate, defined by
\begin{equation}
    \mathrm{SUM}_{c\rightarrow t}
    \coloneqq e^{-i\hat q_c\hat p_t},
\end{equation}
acts as the continuous-variable analog of the CNOT gate.
For example, measuring $\hat{S}_q$ can be implemented by preparing an ancilla $\ket{\bar{+}}$, applying $\mathrm{SUM}_{D\to A}$ from the data mode to the ancilla mode, and measuring the ancilla position quadrature.
The $\hat{S}_p$ measurement is implemented analogously using an ancilla $\ket{\bar{0}}$, $\mathrm{SUM}_{A\to D}^\dagger$, and momentum homodyne detection.

\subsection{Surface-GKP code}
A standard way to further suppress residual logical errors is code concatenation~\cite{gottesman_Stabilizer_1997b}.
By treating each GKP code as an inner-code logical qubit and concatenating it with a DV QEC code as the outer code, GKP-level Pauli errors resulting from misidentified displacement shifts can be further suppressed at the outer-code level. 
In this study, we use the surface-GKP code, where the surface code serves as the outer code.
Following the architecture proposed in Ref.~\cite{noh_Faulttolerant_2020a}, we adopt the rotated surface code geometry~\cite{tomita_Lowdistance_2014a,bombin_Optimal_2007}.

For a distance-$d$ rotated surface-GKP code, the stabilizer group contains $2d^2$ GKP stabilizer generators,
\begin{equation}
     \hat{S}_q^{(k)} \coloneqq e^{i 2 \sqrt{\pi} \hat{q}_k}, \ \hat{S}_p^{(k)} \coloneqq e^{-i 2 \sqrt{\pi} \hat{p}_k },
\end{equation}
for $k \in \{0, \cdots, d^2 - 1\}$. 
It also includes $d^2 - 1$ surface code stabilizer generators:
\begin{equation}
    \begin{gathered}
        \mathcal{S}_{\mathrm{surface}} \coloneqq \langle \hat{S}_Z^{(\ell)}, \hat{S}_X^{(\ell)} \rangle, \\
        \hat{S}_Z^{(\ell)} \coloneqq \prod_{k \in v_Z(\ell)} \bar{Z}_{\mathrm{gkp}}^{(k)}, \ \hat{S}_X^{(\ell)} \coloneqq \prod_{k \in v_X(\ell)} \bar{X}_{\mathrm{gkp}}^{(k)},
    \end{gathered}
\end{equation}
for $\ell \in \{0, \cdots, (d^2 - 3)/2\}$. 
Here, $v_Z(\ell)$ and $v_X(\ell)$ denote the sets of indices for GKP qubits supported by the $\ell$th $Z$-type and $X$-type surface code stabilizer generators, respectively.

Although the Hermitian conjugates of GKP logical operators define the same stabilizer constraints within the GKP codespace, they correspond to different Gaussian gate implementations outside the codespace.
This distinction is important in noisy circuits because displacement errors propagate differently through SUM and inverse-SUM gates.
In Ref.~\cite{noh_Faulttolerant_2020a}, the pattern of SUM and inverse-SUM gates was chosen to mitigate such noise propagation during syndrome extraction.
Specifically, the gate pattern is designed so that propagated shift errors tend to cancel rather than add with the same sign on neighboring syndrome modes, thereby avoiding noise amplification during surface-code syndrome extraction. 
As depicted in Fig.~\ref{fig:surface-gkp}, the surface code syndrome extraction is implemented by coupling the ancilla qubit to the support data qubits via sequential SUM or inverse-SUM gates, followed by a homodyne measurement of the ancilla in the position or momentum basis.

While various decoding strategies have been developed for surface codes~\cite{iolius_Decoding_2024}, we follow the minimum-weight perfect matching~(MWPM)-based decoding approach commonly used in surface-GKP simulations~\cite{noh_Faulttolerant_2020a}.
The MWPM decoder operates on a spacetime syndrome graph constructed from repeated syndrome measurements.
Each node corresponds to a syndrome location, while edges represent possible error processes that connect pairs of syndrome changes across space and time.
Each edge is assigned a weight based on the estimated probability of the corresponding error. In the surface-GKP setting, analog information from GKP stabilizer measurements can be incorporated into these weights to improve decoding accuracy.
 
Let $s_{Z,t}^{(\ell)}\in\{0,1\}$ denote the syndrome of the $\ell$th $Z$-type stabilizer at round $t$.
We define the syndrome difference
\begin{equation}
    d_{Z,t}^{(\ell)}
    =
    s_{Z,t}^{(\ell)}
    \oplus
    s_{Z,t-1}^{(\ell)}.
\end{equation}
A detection event is assigned whenever $d_{Z,t}^{(\ell)}=1$.
The $X$-type syndrome difference is defined analogously.

The resulting detection events form the nodes of the decoding graph, and the MWPM algorithm identifies the most likely error configuration by finding a minimum-weight perfect matching between them~\cite{fowler_Minimum_2015,edmonds_Paths_1965}.

\subsection{Analog information}
Analog information, also referred to as soft information, denotes the additional information contained in continuous-valued measurement outcomes in CV systems. 
In the context of GKP codes, this information enables a quantitative estimate of the posterior probability that a syndrome-based correction results in a logical Pauli error, given the continuous-valued measurement outcome and the noise variance. 

For a single quadrature, a logical Pauli error occurs when the displacement is decoded to an equivalent shift differing by an odd multiple of $\sqrt{\pi}$.
For example, a displacement in the interval $[\sqrt{\pi}/2,3\sqrt{\pi}/2)$ results in a logical Pauli error, whereas a displacement in $[3\sqrt{\pi}/2,5\sqrt{\pi}/2)$ differs only by a stabilizer and is therefore decoded correctly.
This alternating structure determines the logical error probability under Gaussian displacement noise.
To quantify this error probability, we model the shift error by a continuous random variable $\xi \sim \mathcal{N}(0,\sigma^2)$. 
Let $E$ denote the event that the GKP correction results in a logical Pauli error for this quadrature. 
This event corresponds to the actual shift $\xi$ falling into one of the intervals
\begin{align}
\left[\left(2n+\frac12\right)\sqrt{\pi},
      \left(2n+\frac32\right)\sqrt{\pi}\right),
\quad n\in\mathbb Z,
\end{align}
which are centered around odd-integer multiples of $\sqrt{\pi}$.
Therefore, the marginal probability of a logical Pauli error before conditioning on a measurement outcome is
\begin{align}
    \Pr(E; \sigma) 
     &= \frac{1}{\sqrt{2 \pi}\sigma} 
    \sum_{n \in \mathbb{Z}} 
    \int_{(2n+\frac{1}{2})\sqrt{\pi}}^{(2n+\frac{3}{2})\sqrt{\pi}} 
    \exp\left(-\frac{\xi^2}{2\sigma^2}\right) d\xi.
\end{align}

However, if a specific measurement outcome is available, this information can be used to refine the estimate of the logical error probability.
Let $R$ denote the centered modular reduction of $\xi$ modulo $\sqrt{\pi}$, i.e., $R \coloneqq \xi - n\sqrt{\pi}$, where $n$ is the unique integer such that $R \in [-\sqrt{\pi}/2, \sqrt{\pi}/2)$.

Given a measurement outcome $R=r$, the actual shift $\xi$ can differ from $r$ by any integer multiple of $\sqrt{\pi}$, i.e., $\xi=n\sqrt{\pi}+r$ for some $n\in\mathbb{Z}$. 
Among these possibilities, odd values of $n$ correspond to logical Pauli errors. 
Therefore, the conditional probability of a logical Pauli error is
\begin{align}
    \Pr(E\mid R=r;\sigma) = \frac{
    \sum_{n \in \mathbb{Z}} 
    \exp\left(-\frac{((2n+1)\sqrt{\pi}+r)^2}{2\sigma^2}\right)
    }{
    \sum_{n \in \mathbb{Z}} 
    \exp\left(-\frac{(n\sqrt{\pi}+r)^2}{2\sigma^2}\right)
    } .
    \label{eq:analog_info}
\end{align}

\begin{figure}[tb]
    \includegraphics[width=0.97\linewidth]{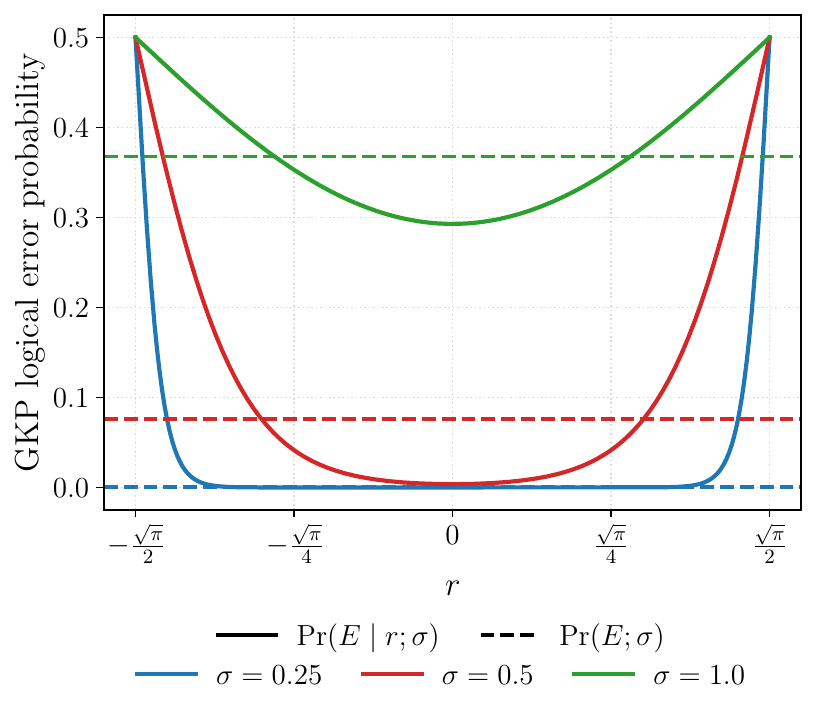}
    \caption{Conditional logical Pauli error probability for a GKP qubit given the measurement outcome $r$ for a single quadrature. 
    The solid line represents the conditional error probability $\Pr(E\mid r;\sigma)$, while the dashed line represents the unconditional error probability $\Pr(E;\sigma)$, which is independent of $r$.}
    \label{fig:analog-info}
\end{figure}

As shown in Eq.~(\ref{eq:analog_info}) and Fig.~\ref{fig:analog-info}, this conditional error probability is minimized when $r$ is near zero and becomes larger as $|r|$ approaches the decision boundary $\sqrt{\pi}/2$. 
Thus, continuous-valued measurement outcomes provide information about the reliability of the corresponding discrete syndrome values.

Previous studies have explored the use of this analog information to enhance code performance~\cite{berent_Analog_2024, fukui_HighThreshold_2018, fukui_Analog_2017}. 
In the specific case of surface-GKP codes, the conditional error probability $\Pr(E\mid r;\sigma)$, or equivalently the reliability $1-\Pr(E\mid r;\sigma)$, can be incorporated into the edge weights of the MWPM decoder, thereby improving decoding accuracy~\cite{noh_Faulttolerant_2020a}. 

In the remainder of this paper, we refer to a logical Pauli error of an inner GKP qubit simply as a Pauli error, since it acts as a physical Pauli error for the outer surface code and should be distinguished from a logical error of the full surface-GKP code.

\section{Adaptive skipping scheme}\label{sec:strategy}
\subsection{Binary-channel model for noisy syndrome extraction}

A central challenge in implementing fault-tolerant QEC is that syndrome extraction is itself noisy. 
In the surface-GKP architecture considered here, a syndrome-extraction circuit involves ancilla GKP-state preparation, gate interactions between data and ancilla qubits, and homodyne measurement, each of which can introduce noise. 
Such noise can both inject additional errors into the data qubits and corrupt the extracted syndrome value. 
In this subsection, we model the latter effect: the possibility that noisy extraction flips the binary syndrome information.
We represent this effect by a binary symmetric channel~(BSC).

\begin{figure}[tb]
    \centering
    \includegraphics[width=0.9\linewidth]{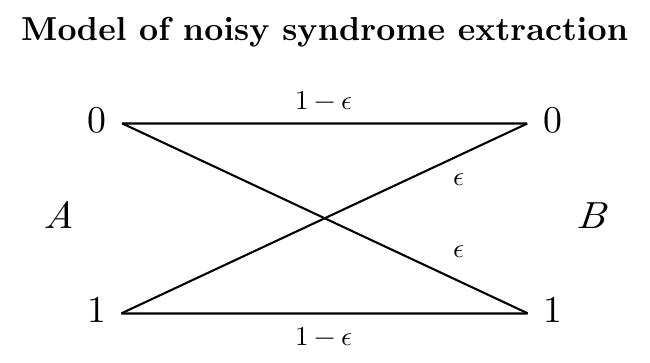}
    \caption{Model of noisy syndrome extraction process as a Binary Symmetric Channel~(BSC). The target syndrome difference $A$ becomes the observed syndrome difference $B$ with flip probability $\epsilon$.}
    \label{fig:BSC-channel}
\end{figure}

In repeated syndrome extraction, MWPM decoding uses detection events, which are determined by changes in syndrome values between consecutive rounds. 
We therefore define the BSC on syndrome differences rather than on the syndrome values themselves, separately for each individual surface-code stabilizer measurement.
For each such stabilizer, let $\hat{s}_t\in\{0,1\}$ denote the syndrome value that would be obtained by noiseless extraction at round $t$, and let $s_{t-1}\in\{0,1\}$ denote the syndrome value recorded in the classical syndrome history from the previous round. The target syndrome difference is then 
\begin{align}
    A:=\hat{s}_t\oplus s_{t-1}.
\end{align}
This recorded value is used rather than the ideal previous syndrome, because the adaptive protocol and the decoder only have access to the recorded syndrome history. Let $\tilde{s}_t\in\{0,1\}$ denote the syndrome value obtained from noisy extraction. The observed syndrome difference is then written as
\begin{align}
    B:=\tilde{s}_t\oplus s_{t-1}.
\end{align}
Noisy syndrome extraction is modeled as a BSC that maps $A$ to $B$ with flip probability $\epsilon:=\Pr(A\neq B)$, as illustrated in Fig.~\ref{fig:BSC-channel}.

We now apply this BSC model to the surface-GKP code.
To understand how the flip probability $\epsilon$ is obtained, we first consider a simplified model in which each qubit in the surface-code experiences independent noise. 
Under this assumption, the noisy extraction of each surface-code stabilizer can be treated as an independent BSC.
For a given surface-code stabilizer measurement, we model the noise on its ancilla GKP qubit as a Gaussian random displacement channel with standard deviation $\sigma_{\mathrm{anc}}$.
If the continuous-valued ancilla measurement outcome $R=r$ were available, the BSC flip probability could be evaluated as $\epsilon = \Pr(E \mid R=r; \sigma_{\mathrm{anc}})$. 
However, in the adaptive decision, this probability must be estimated before performing the corresponding syndrome extraction, so the outcome $r$ is not yet available. 
We therefore use the unconditional flip probability $\epsilon = \Pr(E; \sigma_{\mathrm{anc}})$, obtained by averaging over the unobserved local analog outcome. 

The independent-noise assumption does not hold exactly in the actual circuit-level model. 
In particular, correlated displacement noise introduced by SUM and inverse-SUM gates can create dependencies among different stabilizer measurements.
We nevertheless retain the same BSC structure, using for each stabilizer the marginal standard deviation of its ancilla measurement outcome.
This marginal standard deviation is determined by tracking noise propagation through the surface-GKP syndrome-extraction circuit; the resulting closed-form expressions are given in Appendix~\ref{app:closed}. 
Thus, the BSC flip probability used in the adaptive decision should be understood as a local reliability estimate, rather than as a full joint description of the correlated syndrome-extraction noise.

\subsection{Skip decision rule}
\begin{figure*}[tb]
    \includegraphics[width=\textwidth]{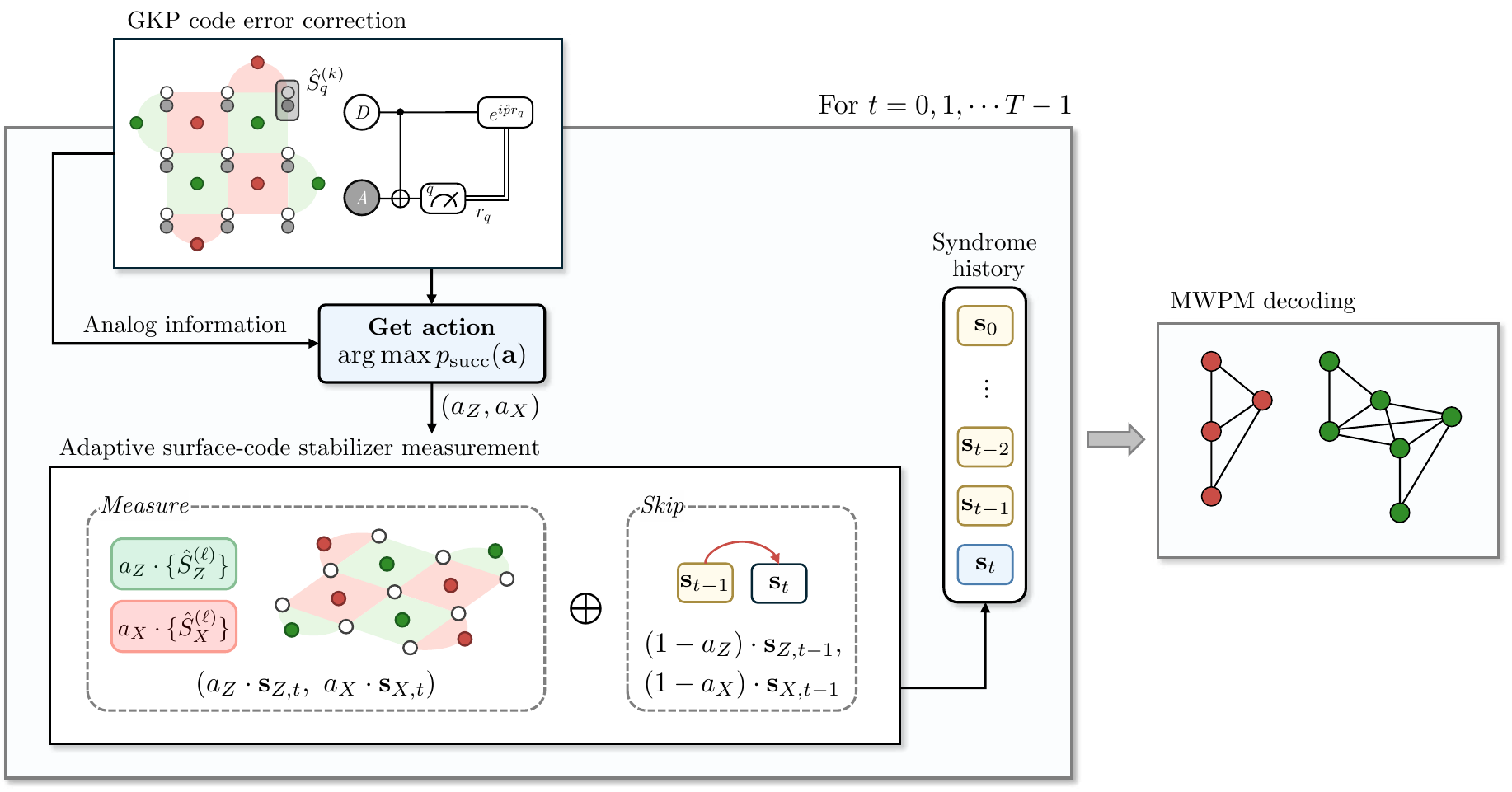}
    \caption{Overview of the adaptive measurement strategy. At each round $t$, GKP stabilizer measurements are always performed to correct small displacement errors and obtain analog information. The protocol then selects an action $\mathbf{a}=(a_Z,a_X)$, where $a_P$ determines whether the $P$-type surface-code stabilizers are measured or skipped. Measured components are obtained from surface-code stabilizer measurements, while skipped components reuse the previous syndrome value. The resulting syndrome history is then used for MWPM decoding after $T$ rounds.}
    \label{fig:overview}
\end{figure*}

The BSC model makes explicit the reliability trade-off behind adaptive syndrome skipping. 
A new syndrome extraction provides fresh information about the current syndrome, but the extracted syndrome difference can be flipped by measurement noise. In addition, the ancilla interactions used for syndrome extraction can inject additional errors into data qubits. 
This motivates a reliability-based rule that compares two alternatives in each round: measuring a new syndrome or skipping the noisy extraction and reusing the previously recorded syndrome value.
Crucially, the reliability of the skip choice is not a fixed prior determined only by the noise model; it is updated in every round from the analog GKP outcomes obtained during inner-code correction.

We refer to the two alternatives for each surface-code stabilizer as \emph{measure} and \emph{skip}.
To make the comparison quantitative, we assign a success probability to each alternative for each surface-code stabilizer. 
Let $P\in\{Z,X\}$ denote the stabilizer type, and let $\hat{S}_P^{(\ell)}$ denote the $\ell$th $P$-type surface-code stabilizer. 
At a given round, we write $A_P^{(\ell)}\in\{0,1\}$ for the target syndrome difference and $B_P^{(\ell)}\in\{0,1\}$ for the syndrome difference obtained from noisy extraction, following the BSC model introduced above. 

We first define the success probability of the \emph{skip} choice for a surface-code stabilizer $\hat{S}_P^{(\ell)}$.
Since skipping syndrome extraction reuses the syndrome value from the previous round, this choice is successful when the ideal syndrome agrees with the previously recorded syndrome value. 
Thus, the skip success probability is defined as
\begin{align}
    p^{\mathrm{skip}}_{\mathrm{succ}}(P;\ell)
    &\coloneqq
    \Pr(A_P^{(\ell)}=0).
\end{align}
To estimate this probability in the adaptive decision rule, we use a local parity approximation. 
Specifically, the previously recorded syndrome value is assumed to be reliable, and the syndrome-flipping Pauli errors on the data qubits involved in the stabilizer are treated as independent after conditioning on the analog information from GKP correction.
Under this approximation, we compute the skip success probability from the analog information obtained during the inner GKP error-correction step,
\begin{align}
    p^{\mathrm{skip}}_{\mathrm{succ}}(P;\ell)
    &\approx
    \frac{1}{2}
    \Big(
        1+\prod_{k\in v_P(\ell)}
        \left(1-2 p_{\mathrm{err}}(P;k)\right) \label{eq:parity}
    \Big),
\end{align}
where $v_P(\ell)$ denotes the set of data-qubit indices in the support of the $\ell$th $P$-type stabilizer, and $p_{\mathrm{err}}(P;k)$ denotes the probability that the $k$th data GKP qubit has a Pauli error that flips a $P$-type stabilizer syndrome.

When the assumptions of the local parity approximation hold exactly, this expression equals the true skip success probability. 
In the circuit-level model considered here, these assumptions are approximate because previous surface-code syndrome-extraction circuits can induce residual correlations among data-qubit shifts. 
We therefore use this expression as a tractable local reliability score for the skip choice.
Since GKP-level Pauli errors are rare events in the low-noise regime, these residual correlations are expected to have a limited effect on the local reliability scores.

The quantities $p_{\mathrm{err}}(P;k)$ are the main channel through which analog GKP information enters the adaptive decision. 
They are round-dependent posterior error probabilities computed from the modular GKP measurement outcomes, rather than fixed physical error rates. 
Since a $Z$-type surface-code syndrome is flipped by an $X$ error, we use the conditional error probability inferred from the $\hat{S}_q^{(k)}$ measurement outcome. 
Similarly, since an $X$-type surface-code syndrome is flipped by a $Z$ error, we use the conditional error probability inferred from the $\hat{S}_p^{(k)}$ measurement outcome:
\begin{equation}
    p_{\mathrm{err}}(P;k)
    =
    \begin{cases}
        \Pr(E_X^{(k)} \mid r_q^{(k)};\sigma_q^{(k)}), & P=Z, \\
        \Pr(E_Z^{(k)} \mid r_p^{(k)};\sigma_p^{(k)}), & P=X.
    \end{cases}
\end{equation}
Here, $E_X^{(k)}$ and $E_Z^{(k)}$ denote the events that the $k$th data GKP qubit has an $X$ and $Z$ error, respectively. 
The quantities $r_q^{(k)}$ and $r_p^{(k)}$ are the corresponding modular inner GKP measurement outcomes, and $\sigma_q^{(k)}$ and $\sigma_p^{(k)}$ are the effective standard deviations of those measurements.

Next, we define the success probability of the \emph{measure} choice for a surface-code stabilizer $\hat{S}_P^{(\ell)}$. 
When syndrome extraction is performed, this choice is successful if the observed syndrome difference agrees with the target syndrome difference, i.e., if $A_P^{(\ell)}=B_P^{(\ell)}$. 
Thus, the measure success probability is defined as
\begin{align}
    p^{\mathrm{meas}}_{\mathrm{succ}}(P;\ell)
    &\coloneqq \Pr(A_P^{(\ell)}=B_P^{(\ell)}).
\end{align}
Within the local BSC model introduced above, the noisy extraction flips the syndrome difference with probability $\epsilon_P^{(\ell)}$.
Therefore,
\begin{align}
    p^{\mathrm{meas}}_{\mathrm{succ}}(P;\ell)
    =
    1-\epsilon_P^{(\ell)}.
\end{align}
At the time of the adaptive decision, the corresponding syndrome extraction has not yet been performed, so its measurement outcome is unavailable.
We therefore evaluate $\epsilon_P^{(\ell)}$ using the unconditional GKP logical-error probability determined by the marginal standard deviation of the corresponding ancilla measurement outcome.
Thus, the two reliability estimates use different information: the measure reliability is a prior estimate based on the expected noise of a surface-code syndrome extraction, whereas the skip reliability is conditioned on the analog GKP outcomes already observed in the current round.

Using the \textit{skip} and \textit{measure} success probabilities defined above, we now construct the global decision rule.
At each round, the adaptive skipping scheme selects one of four actions,
\begin{equation}
    \mathbf{a}=(a_Z,a_X)\in\{0,1\}^2,
\end{equation}
where $a_P=1$ means that all $P$-type surface-code stabilizers are measured, and $a_P=0$ means that all $P$-type stabilizers are skipped. 
Thus, the action $\mathbf{a}$ corresponds to the following measurement choices:
\begin{itemize}
\item $(0,0)$: skip both stabilizer types,
\item $(1,0)$: measure only $Z$-type stabilizers,
\item $(0,1)$: measure only $X$-type stabilizers,
\item $(1,1)$: measure both stabilizer types.
\end{itemize}

For each action $\mathbf{a}$, we combine the corresponding local success probabilities into a decision score. 
In this global decision rule, the measure success probability depends on the action $\mathbf{a}$, since the effective noise channel of a surface-code syndrome ancilla depends on which stabilizer types are measured in the same round. 
We therefore write it as $p_{\mathrm{succ}}^{\mathrm{meas}}(P;\ell\mid \mathbf{a})$. 
The effective standard deviations used to compute 
$p_{\mathrm{succ}}^{\mathrm{meas}}(P;\ell\mid \mathbf{a})$ are derived in Appendix~\ref{app:closed}.

\begin{algorithm}[H]
\caption{Adaptive skipping scheme}
\begin{algorithmic}[1]
\State Initialize the logical state as $\ket{\bar{0}}$
\State Initialize $(\mathbf{s}_{Z,-1},\mathbf{s}_{X,-1}) \leftarrow (\mathbf{0},\mathbf{0})$
\For{$t = 0,\ldots,T-1$}
    \State Prepare ancilla qubits for GKP stabilizer measurements
    \State $\{r_q^{(k)},r_p^{(k)}\}_k
        \leftarrow \textsc{GKPMeasurement}$
    \State $\{\xi_q^{(k)},\xi_p^{(k)}\}_k
        \leftarrow \textsc{GKPErrorCorrection}$
    \State Get action $
        (a_Z^\star,a_X^\star)
        \leftarrow
        \argmax_{\mathbf{a}\in\{0,1\}^2}
        p_{\mathrm{succ}}(\mathbf{a})
        $
    \State Let $\mathcal{P}
        \leftarrow \{P\in\{Z,X\}:a_P^\star=1\}$
    \If{$\mathcal{P} \neq\emptyset$}
        \State Prepare ancilla qubits for the selected surface-code
        \\\hspace{3em}stabilizer measurements
        \State $\{\mathbf{s}_{P,t}^{\mathrm{meas}}:P\in\mathcal{P}\}
            \leftarrow \textsc{SurfaceMeasurement}(\mathcal{P})$
    \Else
        \State Skip all surface-code stabilizer measurements
    \EndIf
    \State For each $P\in\{Z,X\}$, set
    \[
        \mathbf{s}_{P,t}\leftarrow
        \begin{cases}
            \mathbf{s}_{P,t}^{\mathrm{meas}}, & a_P^\star=1,\\
            \mathbf{s}_{P,t-1}, & a_P^\star=0.
        \end{cases}
    \]
\EndFor
\State Find detection events $\{(t,\ell):d_{Z,t}^{(\ell)}=1\}$, 
$\{(t,\ell):d_{X,t}^{(\ell)}=1\}$
\State Construct the decoding graph from all detection events
\State Run MWPM decoding
\State Apply the resulting recovery to the data qubits
\end{algorithmic}
\label{alg:adaptive}
\end{algorithm}

For tractability, we do not compute the full joint success probability directly. 
Instead, we treat the local success events for different stabilizers as conditionally independent given the action $\mathbf{a}$.
Under this independence approximation, we define the decision score as
\begin{align}
    p_{\mathrm{succ}}(\mathbf{a})
    &:=
    \prod_{P\in\{Z,X\}}
    \prod_{\ell}
        \big[
            a_P\, p_{\mathrm{succ}}^{\mathrm{meas}}(P;\ell\mid \mathbf{a})
        \nonumber \\
    &\hspace{2.3cm}
        +
            (1-a_P)\, p_{\mathrm{succ}}^{\mathrm{skip}}(P;\ell)
        \big].
    \label{eq:global_score}
\end{align}
If the local success events were indeed conditionally independent given $\mathbf{a}$, this product would coincide with the joint probability that all local success events occur under action $\mathbf{a}$.
In the circuit-level model considered here, however, these local success events need not be independent.
Measurement-success events can be correlated through gate-noise covariance and noise propagation, while skip-success events can be correlated through shared data qubits and residual correlations among data-qubit Pauli-error indicators.
We therefore use $p_{\mathrm{succ}}(\mathbf{a})$ as a tractable decision score rather than as an exact global posterior success probability.

The adaptive skipping scheme then selects the action with the largest score:
\begin{align}
    \mathbf{a}^\star 
    :=
    \argmax_{\mathbf{a}\in\{0,1\}^2}
    p_{\mathrm{succ}}(\mathbf{a}).
\end{align}

We restrict the adaptive decision to these four global actions rather than making separate decisions for individual stabilizers. 
This restriction keeps the action-dependent noise model tractable. 
If decisions were made separately for individual stabilizers, the measurement success probability of a given stabilizer would depend on which neighboring stabilizers are measured or skipped in the same round, because the propagated noise on its ancilla depends on the local pattern of measured and skipped stabilizers.
Tracking all such pattern-dependent noise channels would substantially complicate the decision rule.
Therefore, we compare only the four candidate global actions using the independence approximation introduced above, while still accounting for the action dependence of the local variances used in the measurement-success probabilities. 
The overall procedure is summarized in Fig.~\ref{fig:overview} and Algorithm~\ref{alg:adaptive}.

\begin{figure}[t]
    \centering
    \includegraphics[width=\linewidth]{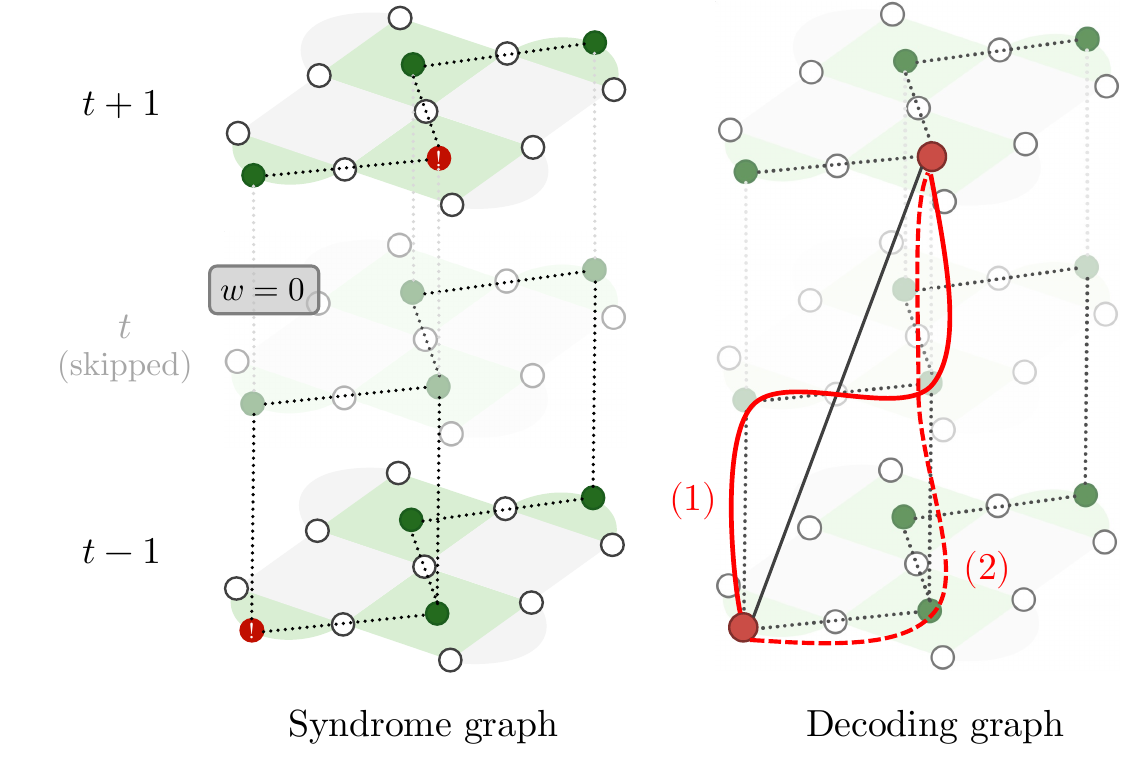}
    \caption{
    Decoding with skipped stabilizer measurements.
    When a round is skipped, the previous syndrome value is reused and no detection event is generated.
    The corresponding time-like edge, shown as the gray dashed line, is retained with weight $w=0$, so paths such as (1) and (2) can pass through the skipped interval without additional penalty.
    }
    \label{fig:decoder}
\end{figure}

\subsection{Decoding}\label{subsec:decoding}
As discussed in Sec.~\ref{sec:background}, surface-GKP decoding is performed by applying MWPM to a three-dimensional decoding graph. 
In the standard setting, the time-like edge weights in the syndrome graph are determined from the reliability of the corresponding surface-code stabilizer measurements. 
In the adaptive skipping scheme, however, some stabilizer measurements are not performed. 
For a skipped stabilizer, no new measurement outcome is generated; instead, the previously recorded syndrome value is reused.
The corresponding time-like edge, therefore, should not be assigned a measurement-error weight based on a nonexistent measurement outcome.

We adopt the following convention: time-like edges associated with skipped stabilizer measurements are assigned zero weight.
This does not remove the edge from the syndrome graph. 
This convention keeps the skipped interval in the syndrome graph, but assigns no additional penalty to passing through it.
Equivalently, the MWPM decoder can connect detection events across a skipped interval without incurring a spurious time-like cost.
Assigning a positive weight to such an interval could bias the decoder away from time-like paths and toward longer space-like paths, even though no new measurement fault was introduced during the skipped round.
Figure~\ref{fig:decoder} illustrates how zero-weight time-like edges are used in decoding. 
The full decoder-weight assignment rule is given in Appendix~\ref{app:decoding}.

\section{Simulation Results}\label{sec:results} 
\subsection{Simulation setup and noise model}
In this section, we present numerical simulations evaluating the performance of the proposed adaptive syndrome-skipping scheme for the surface-GKP code. 
We use a circuit-level Gaussian displacement noise model that includes the following noise sources: 
\begin{itemize}
    \item GKP state preparation,
    \item gate interactions (SUM and inverse-SUM gates),
    \item homodyne measurement,
    \item idle noise.
\end{itemize}

All noise sources are modeled as Gaussian random displacement channels. 
The preparation of each GKP state is modeled as an ideal GKP state followed by a random displacement channel $\mathcal{N}[\sigma_{\mathrm{gkp}}]$, which adds independent shifts $\xi_q,\xi_p \sim \mathcal{N}(0,\sigma_{\mathrm{gkp}}^2)$ to the two quadratures. 
The GKP noise standard deviation $\sigma_{\mathrm{gkp}}$ is related to the squeezing parameter $s_{\mathrm{gkp}}$, measured in dB, by $s_{\mathrm{gkp}} \coloneqq -10\log_{10}(2\sigma_{\mathrm{gkp}}^2)$.

Following Ref.~\cite{noh_Faulttolerant_2020a}, noisy SUM and inverse-SUM gates are modeled as ideal Gaussian gates followed by correlated Gaussian displacement noise, which arises from noise accumulated during the two-mode gate interaction. 
For a gate acting on a control mode $c$ and a target mode $t$, the added shifts satisfy
$(\xi_q^c,\xi_q^t)\sim \mathcal{N}(0,\Sigma_q)$ and $(\xi_p^c,\xi_p^t)\sim \mathcal{N}(0,\Sigma_p)$, 
where $\Sigma_q$ and $\Sigma_p$ are the covariance matrices for the $\hat{q}$ and $\hat{p}$ quadratures, respectively. 
For the SUM gate, these covariance matrices are
\begin{equation}
\Sigma_q=\sigma_{\mathrm{gate}}^2
\begin{pmatrix}
1 & 1/2 \\
1/2 & 4/3
\end{pmatrix}, \quad
\Sigma_p=\sigma_{\mathrm{gate}}^2
\begin{pmatrix}
4/3 & -1/2 \\
-1/2 & 1
\end{pmatrix},
\end{equation}
whereas for the inverse-SUM gate they are
\begin{equation}
\Sigma_q=\sigma_{\mathrm{gate}}^2
\begin{pmatrix}
1 & -1/2 \\
-1/2 & 4/3
\end{pmatrix}, \quad
\Sigma_p=\sigma_{\mathrm{gate}}^2
\begin{pmatrix}
4/3 & 1/2 \\ 
1/2 & 1
\end{pmatrix}.
\end{equation}
Here, $\sigma_{\mathrm{gate}}^2$ denotes the variance scale of the gate-induced displacement noise.

Similarly, noisy homodyne measurement is modeled by applying a random displacement channel 
$\mathcal{N}[\sigma_{\mathrm{meas}}]$ immediately before an ideal homodyne measurement. 
During idle periods, each GKP qubit is subjected to an independent random displacement channel 
$\mathcal{N}[\sigma_{\mathrm{idle}}]$. 
Note that even when a surface-code syndrome measurement is skipped, the data GKP qubits are still exposed to idle noise during the corresponding syndrome-extraction interval.

In our simulations, we use a single circuit-noise parameter $\sigma$ by setting \begin{equation} 
    \sigma^2 \coloneqq \sigma_{\mathrm{gate}}^2 = \sigma_{\mathrm{meas}}^2 = 2\sigma_{\mathrm{idle}}^2 . 
\end{equation} 
This leaves two independent noise parameters, $\sigma$ and $\sigma_{\mathrm{gkp}}$.
The choice $\sigma_{\mathrm{idle}}^2=\sigma^2/2$ reflects regimes in which idle errors are suppressed relative to errors from active circuit operations.
More detailed simulation settings and additional numerical results are provided in Appendices~\ref{app:detail} and~\ref{app:result}.

\begin{figure*}[tb]
    \includegraphics[width=\linewidth]{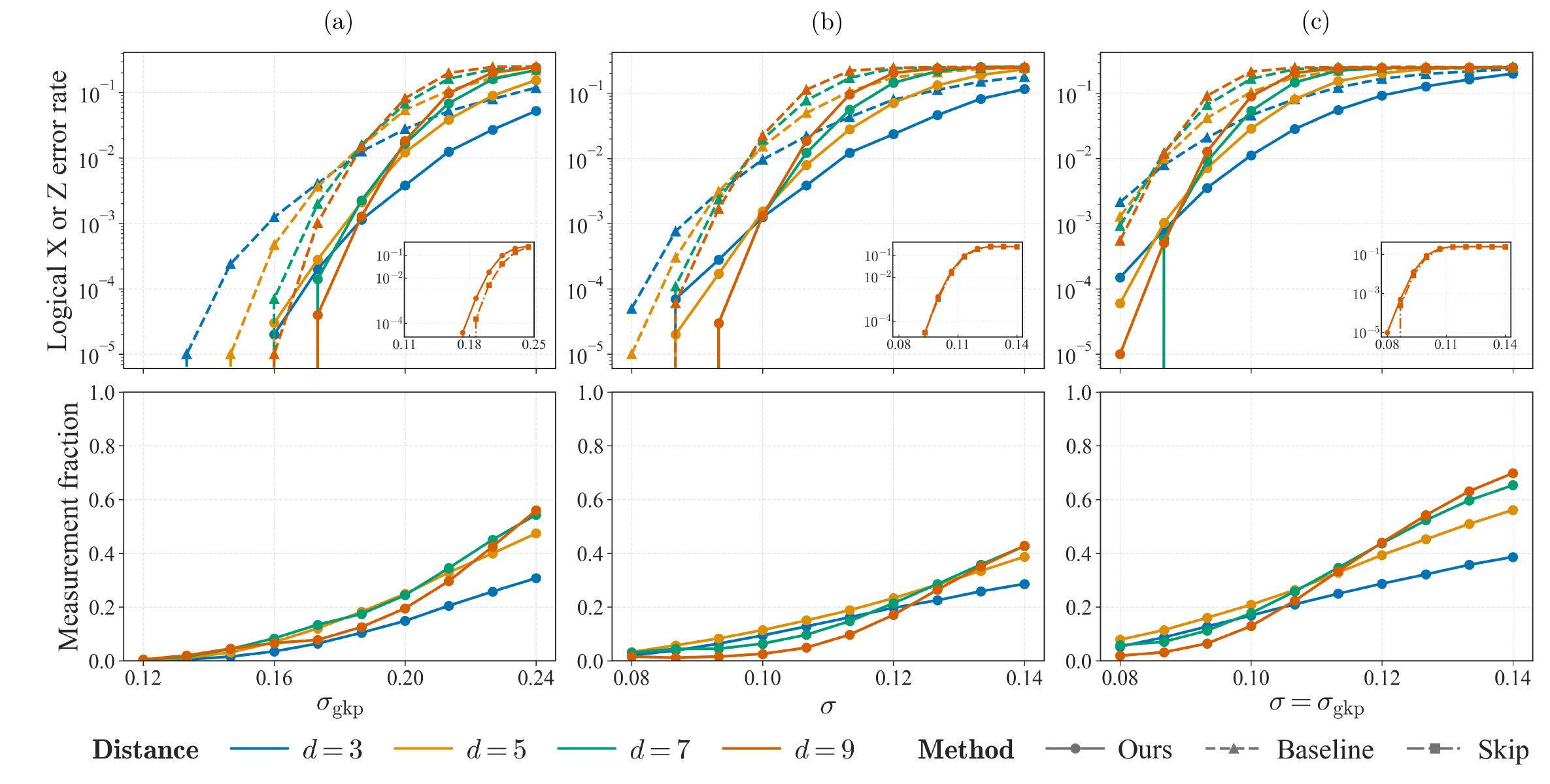}
    \caption{
    Finite-round performance of the adaptive skipping scheme. 
    For each distance-$d$ surface-GKP code, $d$ noisy syndrome-extraction rounds are performed, followed by one final noiseless round. 
    The top panels show the logical $X$ or $Z$ error rate, and the bottom panels show the measurement fraction, defined as the average number of measured surface-code stabilizers normalized by that of the full-measurement baseline. 
    We compare three noise regimes: 
    (a) $\sigma=0$, 
    (b) $\sigma_{\mathrm{gkp}}=0$, and 
    (c) $\sigma=\sigma_{\mathrm{gkp}}$. 
    Solid lines represent the adaptive skipping scheme, dashed lines represent the full-measurement baseline, and the insets compare the adaptive skipping scheme with the all-skip strategy for $d=9$.
    }
    \label{fig:short}
\end{figure*}

\subsection{Finite-round performance}\label{subsec:finite}
We first evaluate the adaptive skipping scheme in the standard finite-round setting used for surface-code simulations. 
For each distance-$d$ surface-GKP code, we perform $d$ noisy rounds of surface-GKP error correction, followed by one final noiseless round, and then apply MWPM decoding to the resulting syndrome history. 
In each noisy round, inner GKP correction is always performed, while outer surface-code stabilizer measurements are either performed or skipped depending on the strategy.
We compare the proposed adaptive skipping scheme with the full-measurement baseline, in which all surface-code stabilizers are measured in every noisy round. 
In both methods, analog information from GKP stabilizer measurements is incorporated into the syndrome graph edge weights.

We consider three representative cases: (a) $\sigma=0$, (b) $\sigma_{\mathrm{gkp}}=0$, and (c) $\sigma=\sigma_{\mathrm{gkp}}$. 
For each case, we simulate distance-$d$ surface-GKP codes with $d=3,5,7,9$, varying the relevant noise parameter for both the baseline and the adaptive skipping scheme.

Figure~\ref{fig:short} shows the logical error rate and the measurement fraction for the three representative noise regimes. 
Here, the measurement fraction is defined as the average number of measured surface-code stabilizers normalized by that of the full-measurement baseline. 
Across all three cases, the adaptive skipping scheme achieves logical error rates comparable to or lower than those of the full-measurement baseline while substantially reducing the number of surface-code stabilizer measurements. 
In the low-noise regime, the logical error rate decreases as the code distance increases, suggesting that larger surface-code distances provide additional error suppression in this regime.
The measurement fraction is also small in this regime and increases with the noise strength. 
This behavior indicates that the adaptive rule skips surface-code syndrome extraction mainly when the analog GKP outcomes suggest that the corresponding surface-code syndrome values are likely to remain unchanged.
In the low-noise regime, the adaptive rule can therefore skip most surface-code stabilizer measurements, making its behavior close to the limiting case in which no surface-code stabilizers are measured during the noisy rounds.

These results highlight the central trade-off underlying this concrete adaptive rule.
A new surface-code stabilizer measurement provides fresh syndrome information, but it is not always beneficial: the noisy extraction circuit can both flip the extracted syndrome value and inject additional errors into the data GKP qubits. 
The adaptive rule uses analog GKP information to identify rounds in which the previous syndrome value is likely to remain reliable, so that the information gained by a new extraction does not justify the additional circuit-induced noise. 
Because $\sigma_{\mathrm{idle}} < \sigma$ in our simulations, skipping such measurements reduces exposure to active syndrome-extraction noise while replacing it only with weaker idle noise. 
This trade-off helps explain why, in the finite-round parameter regimes considered here, this simple reliability-based rule can achieve logical error rates comparable to or lower than those of the full-measurement baseline while using fewer surface-code stabilizer measurements.

This trade-off appears in all three cases.
In case~(a), only the GKP states are noisy, while circuit and idle noise are absent. 
Thus, skipping surface-code syndrome extraction removes the finite-squeezing noise of the GKP ancilla without adding idle noise. 
In case~(b), the GKP states are ideal, while the circuit elements are noisy. 
In this case, the analog information from GKP stabilizer measurements remains relatively reliable, allowing the scheme to avoid unnecessary noisy surface-code syndrome extraction.
In case~(c), both GKP noise and circuit noise are present. This case includes both sources of noise and is therefore the most representative regime. 
Nevertheless, the adaptive skipping scheme achieves logical error rates comparable to, and in some regimes lower than, the full-measurement baseline while using fewer surface-code stabilizer measurements.

The small measurement fractions observed in the low-noise regime motivate a comparison with an all-skip strategy, in which all surface-code stabilizer measurements are skipped in every noisy round.
Note that GKP stabilizer measurements are still performed. 
The insets of Fig.~\ref{fig:short} compare the adaptive skipping scheme and the all-skip strategy for $d=9$. 
As shown in the insets, the all-skip strategy gives performance comparable to, or even better than, the adaptive skipping scheme over much of the parameter range. 
This comparison should be interpreted as a feature of the present finite-round setting and of the particular global reliability rule used here, leaving room for more refined adaptive policies.
This behavior reflects two features of the present finite-round simulation setting. 
First, since idle noise is weaker than circuit noise, skipping reduces the total noise injected into the code, and the errors accumulated on the data qubits during the skipped intervals can often be suppressed by GKP stabilizer measurements alone. 
Second, even when surface-code syndrome extraction is skipped, the MWPM decoder still uses analog information obtained from the GKP stabilizer measurements during the skipped rounds. 
This analog information assigns round-dependent space-like edge weights in the three-dimensional syndrome graph and helps the decoder identify likely GKP-level error locations. If this information is discarded and decoding is performed only using the two-dimensional syndrome graph obtained from the final noiseless surface-code measurement, performance degrades significantly. This comparison is shown in Appendix~\ref{app:finite}, Fig.~\ref{fig:app-skip-analog}.

These observations show that the present simulation setting can make skipping appear especially favorable. 
For this reason, finite-round performance alone does not fully reveal the benefit of making measurement decisions adaptively, motivating the long-time analysis presented in Sec.~\ref{subsec:long}.
When idle noise is comparable to or larger than circuit noise, the advantage of all-skip is reduced, and the all-skip strategy performs worse than the adaptive skipping scheme.  
These results are shown in Appendix~\ref{app:finite}, Figs.~\ref{fig:app-same} and~\ref{fig:app-worse}.

\subsection{Long-time behavior}\label{subsec:long}
Next, we evaluate this concrete adaptive rule in a longer-time setting. 
Instead of decoding immediately after $d$ noisy rounds, we perform decoding after a larger number of noisy rounds. 
This setting more clearly distinguishes the adaptive skipping scheme from the all-skip strategy.
Over a short time interval, errors accumulated on the data GKP qubits can often be suppressed by GKP error correction alone. 
However, this does not necessarily imply that skipping surface-code syndrome extraction remains effective over longer time scales. 
Since the adaptive rule behaves similarly to the all-skip strategy in the low-noise finite-round regime, we examine whether its measurement decisions remain useful over longer time scales, where accumulated errors can make syndrome refreshes more important.
For this purpose, we use the logical error rate per round,
\begin{equation}
    p_L = 1 - \left(1 - p_t\right)^{1/t},
\end{equation}
where $p_{t}$ is the logical error rate after $t$ noisy rounds.

Without periodic full-measurement rounds, both the all-skip strategy and this adaptive rule fail to reach a stable effective logical error rate per round as $t$ increases, unlike the full-measurement baseline; see Appendix~\ref{app:long}, Fig.~\ref{fig:divergence}. 
This behavior reflects a limitation of skipping-based strategies when surface-code syndrome information is not regularly refreshed. 
Therefore, in the following simulations, we introduce a simple refresh mechanism by forcing a full surface-code syndrome extraction every $d$ rounds.
 
We simulate distance-$d$ surface-GKP codes with $d=5,9$ for up to $T=50$ noisy rounds. 
We fix the noise parameters to $\sigma=\sigma_{\mathrm{gkp}}=0.0785$ and $\sigma_{\mathrm{idle}}=0.0555$, a parameter regime close to the threshold-like transition discussed in Sec.~\ref{sec:threshold}.
For the full-measurement baseline at distance $d=9$, we report data up to $T=32$ rounds because its MWPM decoding cost grows rapidly with $T$. 
The available data already place the baseline in the relevant performance range for comparison with this adaptive rule, so extending the baseline simulation to larger $T$ is not needed for the qualitative comparison made here.

\begin{figure}[t]
    \includegraphics[width=\linewidth]{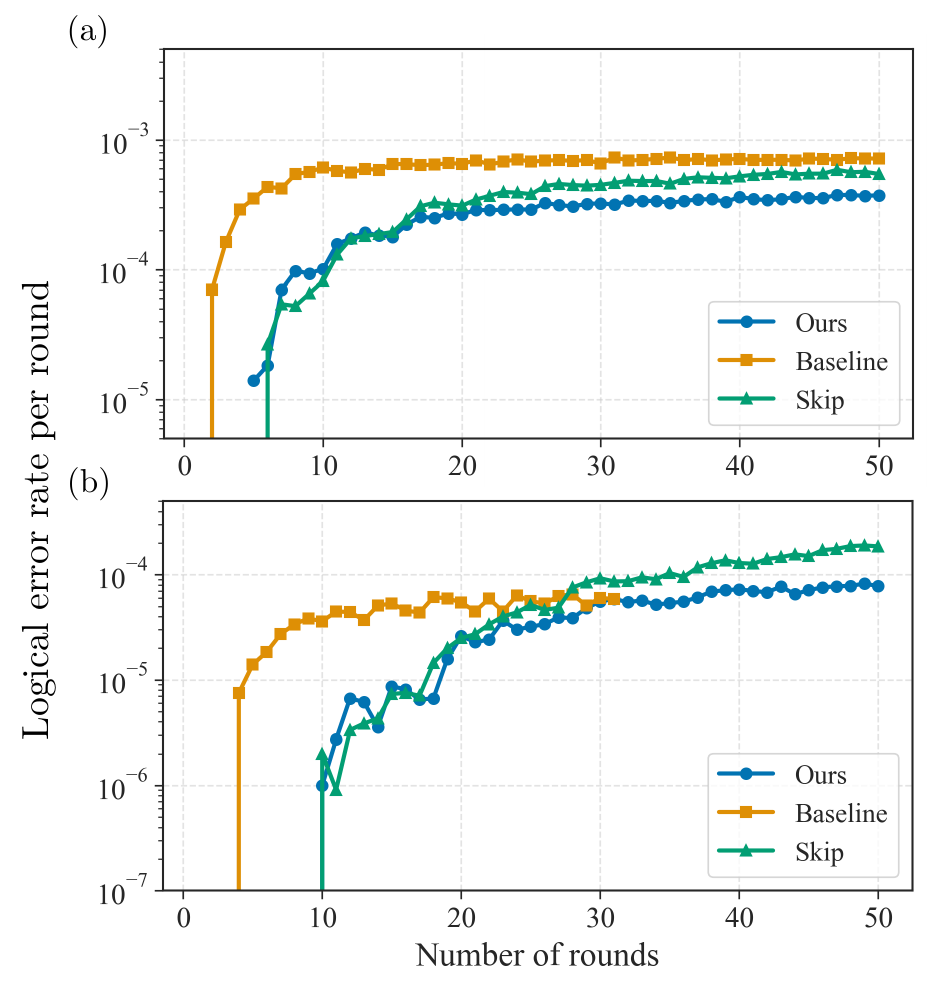}
    \caption{
    Logical error rate per round as a function of the number of noisy rounds for (a) $d=5$ and (b) $d=9$.
    A full surface-code syndrome extraction is forced every $d$ rounds.
    }
    \label{fig:long}
\end{figure}

As shown in Fig.~\ref{fig:long}, forcing a full surface-code syndrome extraction every $d$ rounds makes the effective logical error rate per round approach a stable value for both the all-skip strategy and the adaptive skipping scheme. However, the all-skip strategy remains less stable over long time scales. For both $d=5$ and $d=9$, the all-skip strategy can initially achieve a lower logical error rate per round, but its performance degrades relative to the adaptive scheme as the number of rounds increases. This indicates that periodic full syndrome extraction alone does not fully compensate for the loss of syndrome information during the skipped rounds.

By contrast, the adaptive skipping scheme can perform additional surface-code stabilizer measurements when the stored syndrome values become less reliable, for example, due to error accumulation during skipped intervals or noise introduced by forced full-measurement rounds.
Although it may behave similarly to the all-skip strategy in very low-noise regimes or over short time scales, it does not remain locked into the all-skip behavior when the analog information indicates that new syndrome information is needed. This feedback mechanism improves the long-time stability compared with the all-skip strategy.

In this long-time benchmark, the full-measurement baseline still provides the most stable reference curve. 
This comparison highlights the trade-off inherent in reduced-measurement strategies: skipping surface-code syndrome extraction reduces measurement overhead and avoids some circuit-induced noise, but it also relies on stored syndrome values in the skipped rounds. 
The adaptive rule mitigates this effect relative to the all-skip strategy by performing additional measurements when the analog information indicates that new syndrome information is useful. 
However, with the simple global decision rule used here, this reduced-measurement strategy does not fully recover the long-time stability of continuously refreshing all surface-code syndromes. 
This suggests that more refined adaptive policies may further improve the balance between syndrome refreshes and measurement-induced noise.

\subsection{Distance scaling and threshold-like behavior}\label{sec:threshold}
Finally, we examine the distance scaling of the adaptive skipping rule. 
The finite-round simulations above suggest threshold-like behavior as the noise strength is varied. 
To test whether this behavior persists at larger code distances, we simulate surface-GKP codes up to distance $d=17$, using $d$ noisy rounds of surface-GKP error correction for each distance. 
We use the same noise setting as above: the circuit noise and GKP-state noise are set equal, $\sigma=\sigma_{\mathrm{gkp}}$, and the idle-noise strength is chosen such that $2\sigma_{\mathrm{idle}}^2=\sigma^2$.

\begin{figure}[bt]
    \includegraphics[width=\linewidth]{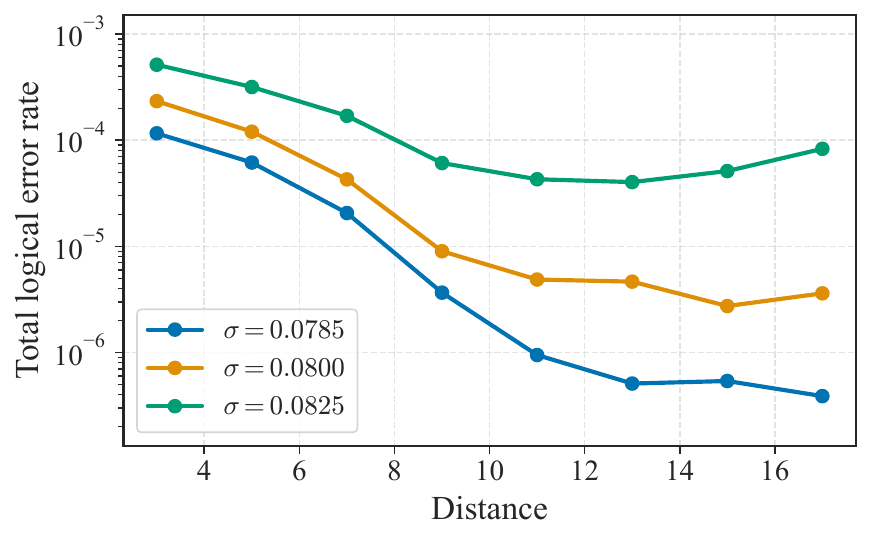}
    \caption{Distance scaling of the adaptive skipping rule. 
    The total logical error rate is plotted as a function of the code distance.
    For each distance-$d$ surface-GKP code, $d$ noisy rounds of surface-GKP error correction are performed, followed by one final noiseless round. 
    }
    \label{fig:thresh}
\end{figure}

Figure~\ref{fig:thresh} shows the total logical error rate as a function of the code distance for several values of $\sigma$. 
Here, the total logical error rate is defined as the sum of the logical $X$, $Y$, and $Z$ error rates. 
The distance dependence changes around $\sigma\simeq 0.08$, suggesting threshold-like behavior near this value. 
For $\sigma>0.08$, the total logical error rate tends to increase with distance, especially at larger distances. 
In contrast, for $\sigma\le 0.08$, it decreases as the code distance increases.  
These results provide further evidence that the adaptive skipping rule can exhibit threshold-like behavior in the circuit-level noise model considered here.

Therefore, unlike the two fixed strategies considered above, namely the full-measurement baseline and the all-skip strategy, the adaptive skipping rule provides an intermediate approach that adapts to the noise environment. 
The rule adjusts its behavior based on the overall noise strength and the relative sizes of circuit and idle noise. 
This makes adaptive skipping particularly relevant to regimes where syndrome extraction itself introduces substantial noise, as may occur in near-term QEC experiments with relatively high physical error rates.

\section{Discussion and Conclusion}\label{sec:discussion_and_conclusion} 
In this work, we have proposed a concrete adaptive skipping scheme for surface-GKP codes. 
The scheme should be viewed as one instantiation of a broader idea: analog information generated by inner GKP correction can be used not only as decoder side information, but also as a control signal for outer-code syndrome extraction. 
At each round, the scheme decides whether to perform selected surface-code stabilizer measurements or reuse the previous syndrome values by comparing two reliability estimates: the probability that the syndrome has not changed from the previous round and the probability that a new noisy syndrome extraction would yield the correct syndrome difference.

Through circuit-level simulations of the surface-GKP code, we have shown that the proposed scheme can reduce the number of surface-code stabilizer measurements while maintaining logical error rates comparable to or lower than those of the full-measurement baseline. 
This behavior is especially relevant in regimes where syndrome extraction itself introduces substantial noise, showing that measuring fewer stabilizers can outperform measuring every stabilizer in every round. 
The long-time simulations further show that periodic refreshes of the surface-code syndrome information are important for stabilizing skipping-based strategies.
Although the adaptive scheme is more stable than the all-skip strategy under periodic full extraction, its stability remains below that of the full-measurement baseline.
This gap suggests that adaptive rules that more accurately track the long-time accumulation of syndrome uncertainty could further improve the stability of syndrome-skipping strategies.

The proposed scheme is also motivated by a practical feature of bosonic QEC architectures. 
In the surface-GKP code, each surface-code stabilizer measurement consumes GKP ancilla states, whose high-fidelity preparation remains experimentally challenging. 
By skipping syndrome measurements that are estimated to be less useful, the adaptive skipping scheme can significantly reduce the number of GKP ancilla states required. 
Thus, the scheme provides a way to reduce resource overhead without changing the underlying code architecture.

Although we have focused on the surface-GKP code, the main idea is not specific to the surface code. 
More generally, this idea can be extended to concatenated architectures that use GKP qubits as an inner code, since analog information from GKP stabilizer measurements can be used to estimate syndrome reliability at the outer-code level. 
We expect that similar adaptive skipping schemes can be applied to other GKP-concatenated codes.

This work represents an initial step toward adaptive syndrome extraction in CV-DV concatenated QEC architectures. 
Several directions remain open for improving the adaptive decision rule. 
For example, one could move from the present global decision rule to a local rule that decides independently for each stabilizer. 
One could also replace the simple comparison of success probabilities with a more refined criterion, such as one based on worst-case or average-case error probabilities. 
In addition, the current rule does not explicitly account for the extra data-qubit errors introduced by noisy syndrome extraction. 
A refined rule that includes this effect could compare skipping and measuring more directly by considering both syndrome reliability and the additional noise injected into the data qubits. 
Such improvements may lead to adaptive protocols that further reduce syndrome-extraction overhead while preserving, or even improving, logical error suppression.

\begin{acknowledgements} 
We would like to thank Kyungjoo Noh for the fruitful discussion.
This work was supported by the National Research Foundation of Korea Grants (No. RS-2024-00431768 and No. RS-2025-00515456) funded by the Korean government (Ministry of Science and ICT (MSIT)) and the Institute of Information \& Communications Technology Planning \& Evaluation (IITP) Grants funded by the Korean government (MSIT) (No. RS-2024-00437284, No. IITP-2025-RS-2025-02283189 and No. IITP-2025-RS-2025-02263264) by Global Partnership Program of Leading Universities in Quantum Science and Technology (RS-2025-08542968) through the National Research Foundation of Korea~(NRF) funded by the Korean government (Ministry of Science and ICT(MSIT)), and by Samsung Electronics Co., Ltd (IO251212-14452-01).
\end{acknowledgements}

\appendix 
\section{Simulation details}\label{app:detail}

The code used for the numerical simulations in this work is publicly available at \url{https://github.com/webb-c/adaptive-surface-GKP}. In the numerical simulations, all logical error rates are estimated by Monte Carlo sampling. For each data point, we collect between $10^4$ and $2\times 10^7$ samples, with the number of samples chosen so that the resulting statistical uncertainties are sufficiently small.

\subsection{Adaptive surface-GKP syndrome-extraction round}

In this section, we describe the simulation procedure for one adaptive surface-GKP syndrome-extraction round. 
Our circuit-level noise model follows Ref.~\cite{noh_Faulttolerant_2020a}, except for two changes. 
First, we introduce an independent idle-noise parameter $\sigma_{\mathrm{idle}}$. 
This allows the relative strength of idle and circuit noise to be varied independently, enabling a more realistic noise model. 
The idle-noise channel $\mathcal{N}[\sigma_{\mathrm{idle}}]$ is applied to every qubit that does not participate in an active operation during a given time step. 
In particular, idle noise is applied in the following cases:
\begin{itemize}
    \item to data qubits during GKP-ancilla preparation,
    \item to data or ancilla qubits that do not participate in a SUM or inverse-SUM gate operation,
    \item to data qubits during homodyne measurements of ancilla qubits.
\end{itemize}
Second, the adaptive decision is applied only to the surface-code syndrome-extraction stage. 
Each round consists of GKP stabilizer measurements followed by surface-code syndrome extraction. 
The GKP stabilizer measurements are always performed, providing both GKP error correction and the analog information used by the adaptive decision rule. 
The selected action then determines whether the $Z$-type and $X$-type surface-code stabilizers are measured or skipped. 

\begin{figure}
    \centering
    \includegraphics[width=\linewidth]{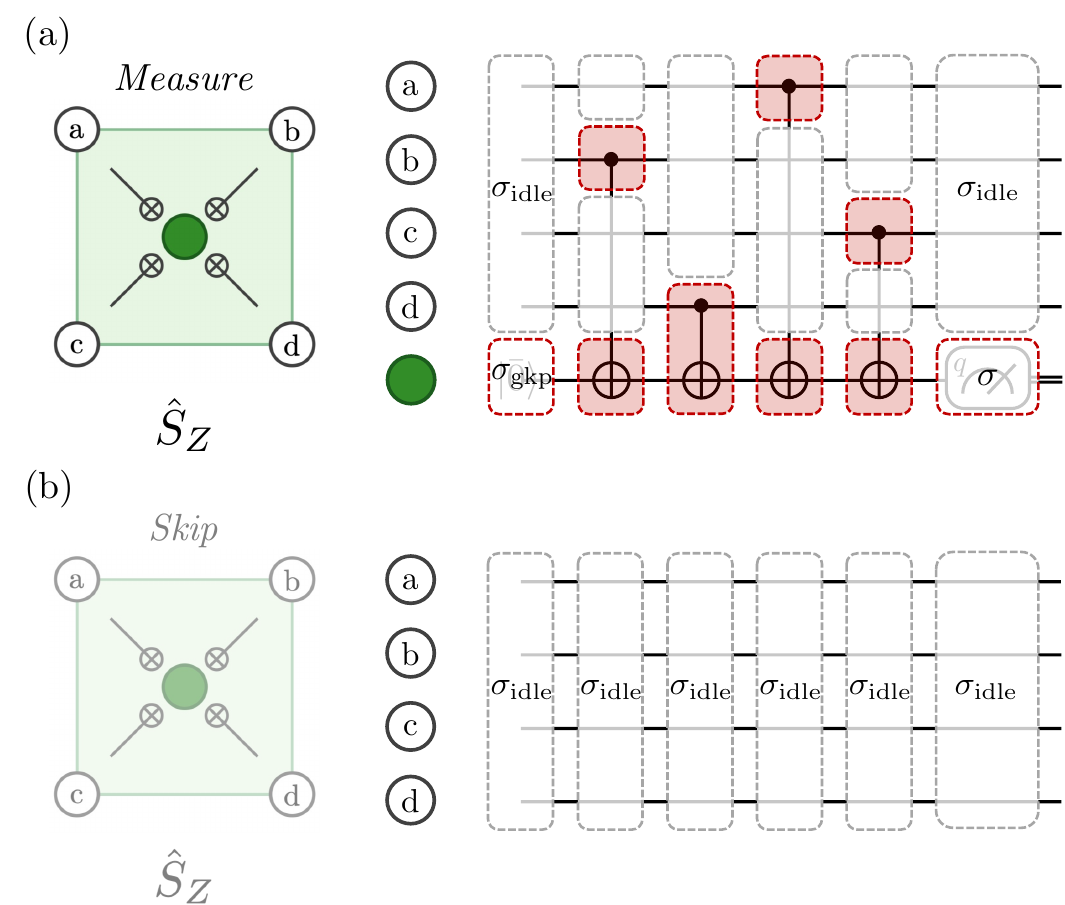}
    \caption{
    Noise channels applied when a $Z$-type surface-code stabilizer measurement is performed or skipped.
    Even when the $Z$-type stabilizer measurement is skipped, the data qubits are exposed to idle noise during the time interval originally allocated to surface-code syndrome extraction.
    Each dashed box represents a noise channel, and $\sigma$ denotes its standard deviation.
    Gray boxes indicate idle-noise channels, while the other boxes indicate non-idle noise components.
    In particular, red boxes denote correlated noise channels arising from gate interactions.
    }
    \label{fig:noise-detail}
\end{figure}

We now clarify what we mean by \emph{skipping} a surface-code stabilizer measurement. 
In our simulations, skipping does not mean that the data qubits are frozen without noise. 
Rather, the data qubits remain idle for the duration in which the corresponding surface-code syndrome-extraction circuit would have been performed. 
Since the surface-code syndrome-extraction circuit consists of ancilla preparation, four sequential gate operations, and homodyne measurement, each data qubit experiences six idle-noise channels during a skipped surface-code syndrome-extraction interval. 
Figure~\ref{fig:noise-detail} illustrates the corresponding noise channels for a representative $Z$-type stabilizer, comparing the measure and skip cases.

\subsection{Noise channel tracking}\label{app:closed}
Since the proposed adaptive skipping scheme is restricted to four global actions,
\begin{equation}
 a\in\{\actNone,\actZ,\actX,\actB\}
 =\{\texttt{skip},\texttt{z\_only},\texttt{x\_only},\texttt{both}\},
\end{equation}
the effective noise standard deviations for all possible cases can be written in closed form.
For compactness, in this appendix, we write an action $a=(a_Z,a_X)$ as $a_Za_X$, where the first and second bits indicate whether the $Z$-type and $X$-type stabilizers are measured, respectively.
These expressions are obtained by composing the Gaussian random-displacement channels applied during the corresponding syndrome-extraction steps.

For a distance-$d$ surface-GKP code, we label the data GKP qubits by
\begin{equation}
 k\in\{0,\ldots,d^{2}-1\},\quad k=id+j, \quad i, j\in\{0,\ldots,d-1\},
\end{equation}
where $i$ and $j$ denote the row and column indices, respectively. 
For the first round, where there is no previous adaptive action, we denote the previous action by $a=\varnothing$. 

Let $\sigma^A_q(a,k)$ denote the effective standard deviation associated with the ancilla qubit used to measure the $q$-quadrature GKP stabilizer $\hat{S}_q^{(k)}$ on the $k$th data qubit, conditioned on the action $a$ chosen in the \emph{previous} round. 
The corresponding quantity for the $p$-quadrature GKP stabilizer measurement is denoted by $\sigma^A_p(a,k)$. 
For the first round, these standard deviations are 
\begin{align}
\sigma^A_q(\varnothing, k) = 
    \begin{cases} 
    \sqrt{\g + \frac{7}{3} \c + \h}, & k\in2\ZZ, \\\
    \sqrt{\g + \frac{11}{3} \c + 3\h},& k\in2\ZZ+1,
    \end{cases}
\\
\sigma^A_p(\varnothing, k) =
    \begin{cases}
    \sqrt{\g + \frac{11}{3} \c + 3\h}, & k\in2\ZZ, \\
    \sqrt{\g + \frac{7}{3} \c + \h},&k\in2\ZZ+1.
    \end{cases}
\end{align}
For later rounds, the corresponding standard deviations depend on the previous action. 
For each possible previous action, they are given as follows. 
\begin{widetext}
For \texttt{skip},
\begin{align}
\sigma^A_q({\actNone},k) = \sqrt{3\g + 6\c + 10\h}, \\ \sigma^A_p({\actNone},k) = \sqrt{3\g + 6\c + 10\h}.
\end{align}
For \texttt{z\_only},
\begin{equation}
\sigma_q^A (\actZ, k) =
\begin{cases}
\sqrt{3 \g + 7\c + 9\h},&i\in\{0,d-1\},\\
\sqrt{3 \g + 8\c + 8\h},&\text{otherwise},
\end{cases}
\end{equation}
and
\begin{equation}
\sigma^A_p(\actZ, k) =
\begin{cases}
\sqrt{4 \g + \frac{22}{3} \c + 9 \h}, &i=0,\ j\in2\ZZ,\\
\sqrt{4 \g + \frac{28}{3} \c + 9 \h}, &i=0,\ j\in2\ZZ+1,\\
\sqrt{5 \g + \frac{29}{3} \c + 10 \h},&0<i<d-1,\ j=d-1,\\
\sqrt{5 \g + \frac{35}{3} \c + 8 \h}, &0<i<d-1,\ j\ne d-1,\\
\sqrt{4 \g + \frac{25}{3} \c + 11 \h},&i=d-1,\ j=d-1,\\
\sqrt{4 \g + \frac{31}{3} \c + 9 \h}, &i=d-1,\ j\in2\ZZ, \ j \ne d-1\\
\sqrt{4 \g + \frac{25}{3} \c + 9 \h}, &i=d-1,\ j\in2\ZZ+1.
\end{cases}
\end{equation}
For \texttt{x\_only},
\begin{equation}
\sigma^A_q(\actX, k) =
\begin{cases}
\sqrt{4 \g + \frac{25}{3} \c + 9 \h}, &j=0,\ i\in2\ZZ,\\
\sqrt{4 \g + \frac{31}{3} \c + 9 \h}, &j=0,\ i\in2\ZZ+1,\\
\sqrt{5 \g + \frac{29}{3} \c + 10 \h},&0 < j < d-1, \ i = 0 \\
\sqrt{5 \g + \frac{35}{3} \c + 8 \h}, &0 < j < d-1, \ i \ne 0\\
\sqrt{4 \g + \frac{22}{3} \c + 11 \h},&j=d-1,\ i = 0,\\
\sqrt{4 \g + \frac{28}{3} \c + 9 \h}, &j=d-1,\ i \in 2 \mathbb{Z},\ i \ne 0\\
\sqrt{4 \g + \frac{22}{3} \c + 9 \h}, &j=d-1,\ i \in 2 \mathbb{Z} + 1,
\end{cases}
\end{equation}
and
\begin{equation}
\sigma_p^A (\actX, k) =
\begin{cases}
\sqrt{3 \g + 7\c + 9\h},&j\in\{0,d-1\},\\
\sqrt{3 \g + 8\c + 8\h},&\text{otherwise}.
\end{cases}
\end{equation}
For \texttt{both},
\begin{equation}
\sigma_q^{A}(\actB,k)=
\begin{cases}
\sqrt{4 \g + \frac{28}{3} \c + 8 \h},&j=0,\ i\in\{0,d-1\},\\
\sqrt{4 \g + \frac{31}{3} \c + 7 \h},&j=0,\ 0<i<d-1,\ i\in2\ZZ,\\
\sqrt{4 \g + \frac{37}{3} \c + 7 \h},&j=0,\ 0<i<d-1,\ i\in2\ZZ+1,\\
\sqrt{5 \g + \frac{32}{3} \c + 9 \h},&0<j<d-1,\ i=0,\\
\sqrt{5 \g + \frac{41}{3} \c + 6 \h},&0<j<d-1,\ 0<i<d-1\\ 
\sqrt{5 \g + \frac{38}{3} \c + 7 \h},&0<j<d-1,\ i=d-1,\\
\sqrt{4 \g + \frac{25}{3} \c + 10 \h},&j=d-1,\ i=0,\\
\sqrt{4 \g + \frac{34}{3} \c + 7 \h},&j=d-1,\ 0<i<d-1,\ i\in2\ZZ,\\
\sqrt{4 \g + \frac{28}{3} \c + 7 \h},&j=d-1,\ 0<i<d-1,\ i\in2\ZZ+1,\\
\sqrt{4 \g + \frac{31}{3} \c + 8 \h},&j=d-1,\ i=d-1,
\end{cases}
\end{equation}
and
\begin{equation}
\sigma_p^{A}(\actB,k)=
\begin{cases}
\sqrt{4 \g + \frac{25}{3} \c + 8 \h},&i=0,\ j\in\{0,d-1\},\\
\sqrt{4 \g + \frac{28}{3} \c + 7 \h},&i=0,\ 0<j<d-1,\ j\in2\ZZ,\\
\sqrt{4 \g + \frac{34}{3} \c + 7 \h},&i=0,\ 0<j<d-1,\ j\in2\ZZ+1,\\
\sqrt{5 \g + \frac{38}{3} \c + 7 \h},&0<i<d-1,\ j=0,\\
\sqrt{5 \g + \frac{41}{3} \c + 6 \h},&0<i<d-1,\ 0<j<d-1,\\
\sqrt{5 \g + \frac{32}{3} \c + 9 \h},&0<i<d-1,\ j=d-1,\\
\sqrt{4 \g + \frac{34}{3} \c + 8 \h},&i=d-1,\ j=0,\\
\sqrt{4 \g + \frac{37}{3} \c + 7 \h},&i=d-1,\ 0<j<d-1,\ j\in2\ZZ, \\
\sqrt{4 \g + \frac{31}{3} \c + 7 \h},&i=d-1,\ 0<j<d-1,\ j\in2\ZZ+1,\\
\sqrt{4 \g + \frac{28}{3} \c + 10 \h},&i=d-1,\ j=d-1.
\end{cases}
\end{equation}
\end{widetext}
These quantities are used in two places: first, to estimate the skip-success probability, and second, to assign the space-like edge weights in the syndrome graph. 
Further details on the decoding procedure are provided in Appendix~\ref{app:decoding}. 

After the GKP error-correction step, the effective standard deviations of the data qubits always take the following values, independently of the previous action:
\begin{align}
\sigma^D_q(k)=
\begin{cases}
\sqrt{2 \g + \frac{11}{3} \c + 3 \h},&k\in2\ZZ,\\
\sqrt{ \g + \frac{7}{3} \c + \h},&k\in2\ZZ+1,
\end{cases}
\\
\sigma^D_p(k)=
\begin{cases}
\sqrt{ \g + \frac{7}{3} \c + \h},&k\in2\ZZ,\\
\sqrt{2 \g + \frac{11}{3} \c + 3 \h},&k\in2\ZZ+1,
\end{cases}
\end{align}
where $\sigma^D_q(k)$ and $\sigma^D_p(k)$ denote the effective standard deviations of the $k$th data qubit in the $q$ and $p$ quadratures, respectively.
Since these values are independent of the previous actions, we can calculate the effective standard deviations in closed form for the corresponding syndrome-extraction steps.

\begin{figure*}
    \centering
    \includegraphics[width=\linewidth]{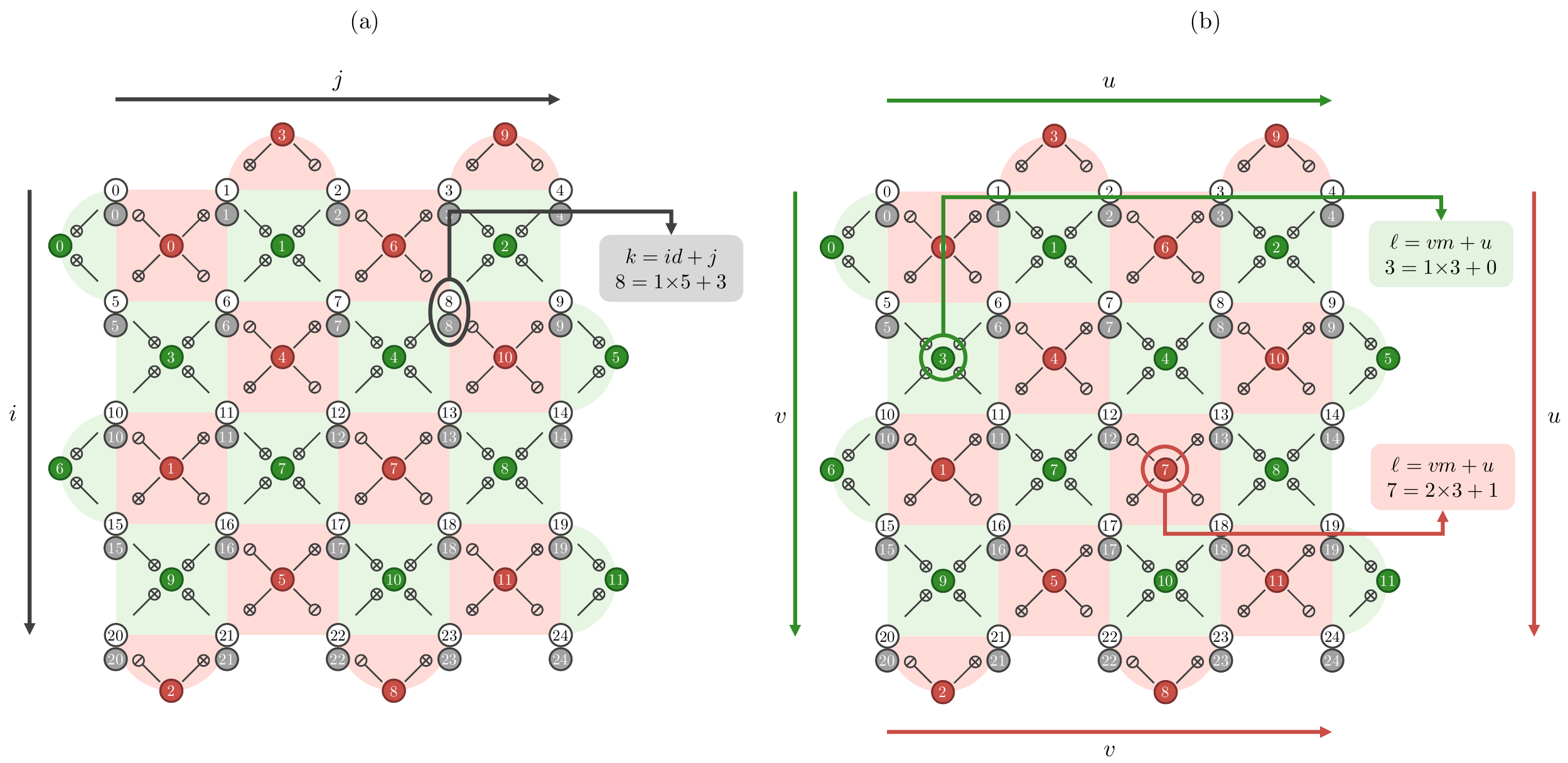}
    \caption{Indexing conventions used in the simulation for the distance-$d=5$ surface-GKP code. White circles denote data qubits, and colored circles denote surface-code syndrome qubits. A data qubit is labeled by the two-dimensional index $(i,j)$, where $i$ and $j$ denote the row and column indices, respectively, and is flattened as $k=id+j$. For example, the data qubit at $(i,j)=(1,3)$ has index $k=1\times 5+3=8$. Surface-code stabilizers are labeled by the two-dimensional index $(u,v)$ and flattened as $\ell=vm+u$, where $m=(d+1)/2$. For $d=5$, we have $m=3$. The green and red annotations show examples of this convention for $Z$-type and $X$-type stabilizers, respectively.
    }
    \label{fig:index}
\end{figure*}

We next give the effective standard deviations for the ancilla qubits used in surface-code stabilizer measurements. 
Unlike the GKP-stabilizer ancilla standard deviations above, these quantities depend only on the current action.
For a distance-$d$ surface-GKP code, we label the surface-code stabilizers by
\begin{equation}
    u=\ell\bmod m,\quad
    v=\left\lfloor\frac{\ell}{m}\right\rfloor,
\end{equation}
where $\ell\in\{0,\ldots,(d^{2}-3)/2\}$ and $m=(d+1)/2$. Here, $u$ and $v$ denote the column and row indices, respectively, of the $\ell$th surface-code stabilizer in our enumeration.

Let $\sigma_Z^A(a,\ell)$ denote the effective standard deviation associated with the ancilla qubit used to measure the $\ell$th $Z$-type surface-code stabilizer $\hat{S}_Z^{(\ell)}$, conditioned on the \emph{current} action $a$.
The corresponding quantity for the $X$-type surface-code stabilizer measurement is denoted by $\sigma_X^A(a,\ell)$. 
Then, for each possible current action, these standard deviations are given as follows.
\begin{widetext}
For \texttt{z\_only},
\begin{equation}
\sigma_Z^A(\actZ,\ell)=
\begin{cases}
\sqrt{4 \g + \frac{29}{3} \c + 9 \h},&u=0,\ v\in2\ZZ,\\
\sqrt{4 \g + \frac{32}{3} \c + 12 \h},&u=m-1,\ v=d-2,\\
\sqrt{4 \g + \frac{35}{3} \c + 11 \h},&u=m-1,\ v\in2\ZZ+1,\ v\ne d-2,\\
\sqrt{7 \g + \frac{58}{3} \c + 17 \h},&v\in\{0,d-2\}\ \text{and not above},\\
\sqrt{7 \g + \frac{61}{3} \c + 16 \h},&\text{otherwise}.
\end{cases}
\end{equation}
For \texttt{x\_only},
\begin{equation}
\sigma_X^A(\actX,\ell)=
\begin{cases}
\sqrt{4 \g + \frac{29}{3} \c + 9 \h},&u=m-1,\ v\in2\ZZ,\\
\sqrt{4 \g + \frac{32}{3} \c + 12 \h},&u=0,\ v=d-2,\\
\sqrt{4 \g + \frac{35}{3} \c + 11 \h},&u=0,\ v\in2\ZZ+1,\ v\ne d-2,\\
\sqrt{7 \g + \frac{58}{3} \c + 17 \h},&v\in\{0,d-2\}\ \text{and not above},\\
\sqrt{7 \g + \frac{61}{3} \c + 16 \h},&\text{otherwise}.
\end{cases}
\end{equation}
For \texttt{both},
\begin{equation}
\sigma_Z^A(\actB,\ell)=
\begin{cases}
\sqrt{4 \g + \frac{29}{3} \c + 9 \h},&u=0,\ v\in2\ZZ,\\
\sqrt{4 \g + \frac{43}{3} \c + 10 \h},&u=m-1,\ v=d-2,\\
\sqrt{4 \g + \frac{46}{3} \c + 9 \h},&u=m-1,\ v\in2\ZZ+1,\ v\ne d-2,\\
\sqrt{7 \g + 22 \c + 15 \h},&u=0,\ v=d-2,\\
\sqrt{7 \g + 23 \c + 14 \h},&u=0,\ v\in2\ZZ+1,\ v\ne d-2,\\
\sqrt{7 \g + \frac{77}{3} \c + 13 \h},&v\in\{0,d-2\}\ \text{and not above},\\
\sqrt{7 \g + \frac{80}{3} \c + 12 \h},&\text{otherwise},
\end{cases}
\end{equation}
and
\begin{equation}
\sigma_X^A(\actB,\ell)=
\begin{cases}
\sqrt{4 \g + \frac{29}{3} \c + 9 \h},&u=m-1,\ v\in2\ZZ,\\
\sqrt{4 \g + \frac{43}{3} \c + 10 \h},&u=0,\ v=d-2,\\
\sqrt{4 \g + \frac{46}{3} \c + 9 \h},&u=0,\ v\in2\ZZ+1,\ v\ne d-2,\\
\sqrt{7 \g + 22 \c + 15 \h},&u=m-1,\ v=d-2,\\
\sqrt{7 \g + 23 \c + 14 \h},&u=m-1,\ v\in2\ZZ+1,\ v\ne d-2,\\
\sqrt{7 \g + \frac{77}{3} \c + 13 \h},&v\in\{0,d-2\}\ \text{and not above},\\
\sqrt{7 \g + \frac{80}{3} \c + 12 \h}&\text{otherwise}.
\end{cases}
\end{equation}
\end{widetext}
Similarly, these quantities are used to estimate the measurement-success probability and to assign the time-like edge weights in the syndrome graph.
Further details on the decoding procedure are provided in Appendix~\ref{app:decoding}.

\subsection{Decoding}\label{app:decoding}
After $T$ noisy syndrome-extraction rounds and one final noiseless data-qubit readout, we obtain a syndrome history consisting of $T+1$ rounds.
From this syndrome history, we construct three-dimensional syndrome graphs and the corresponding decoding graphs as follows. 
The nodes of the decoding graph correspond to detection events in the syndrome graph. 
For each pair of detection events, we compute the shortest-path distance between them in the syndrome graph and use this value as the edge weight in the decoding graph. 
The weights in the syndrome graph are computed from the effective standard deviations of the noise channels derived in Appendix~\ref{app:closed}, together with the corresponding measurement outcomes.

Let $w_Z^H(a,k)$ and $w_X^H(a,k)$ denote the space-like edge weights in the $Z$-type and $X$-type syndrome graphs, respectively, associated with the $k$th data qubit when the previous action was $a$.
Here, the superscript $H$ denotes horizontal space-like edges, and $V$ denotes vertical time-like edges. 
The space-like edge weights are assigned as
\begin{widetext}
\begin{equation}
    w_{Z}^H(a, k) = 
    \begin{cases}
        - \log_2 \Big(\Pr\big(E \mid r_q^{(k)} ; \sigma_q^A(\varnothing, k)\big)\Big), & t = 0, \\
        - \log_2 \Big(\Pr\big(E \mid r_q^{(k)} ; \sigma_q^A(a, k)\big)\Big), & 0 < t < T-1, \\
        - \log_2 \Big(\Pr\big(E \mid r_q^{(k)} ; 
        \sqrt{\big(\sigma_q^A(a, k)\big)^2 - \big(\sigma_q^A(\varnothing, k)\big)^2}
        \big)\Big), & t = T-1,
    \end{cases}
\end{equation}
and
\begin{equation}
    w_{X}^H(a, k) = 
    \begin{cases}
        - \log_2 \Big(\Pr\big(E \mid r_p^{(k)} ; \sigma_p^A(\varnothing, k)\big)\Big), & t = 0, \\
        - \log_2 \Big(\Pr\big(E \mid r_p^{(k)} ; \sigma_p^A(a, k)\big)\Big), & 0 < t < T-1, \\
        - \log_2 \Big(\Pr\big(E \mid r_p^{(k)} ; 
        \sqrt{\big(\sigma_p^A(a, k)\big)^2 - \big(\sigma_p^A(\varnothing, k)\big)^2}
        \big)\Big), & t = T-1.
    \end{cases}
\end{equation}
\end{widetext}
Here, $t$ denotes the round index, and $r_q^{(k)}$ and $r_p^{(k)}$ denote the measurement outcomes of $\hat S_q^{(k)}$ and $\hat S_p^{(k)}$, respectively.

In the adaptive skipping scheme, some surface-code syndrome measurements may be skipped.
For skipped surface-code syndrome measurements, we set the corresponding time-like edge weight in the syndrome graph to zero. 
This zero-weight assignment does not remove the edge. 
Since the other edge weights are nonnegative costs of the form $-\log_2 P$, assigning zero weight gives no additional measurement-error penalty to a syndrome value that was not actually measured. 
This convention is analogous to the zero-weight assignment for virtual boundary edges, in which the edge remains in the syndrome graph but contributes no additional path cost.

Let $w_Z^V(a,\ell)$ and $w_X^V(a,\ell)$ denote the time-like edge weights in the $Z$-type and $X$-type syndrome graphs, respectively, associated with the $\ell$-th surface-code stabilizer when the current action was $a$.
The time-like edge weights are assigned as
\begin{equation}
    w_Z^V(a, \ell) = 
    \begin{cases}
        -\log_2\Big( \Pr \big( E \mid r_Z^{(\ell)} ; \sigma_Z^A(\actZ, \ell) \big) \Big), & a = \actZ,\\
        -\log_2\Big( \Pr \big( E \mid r_Z^{(\ell)} ; \sigma_Z^A(\actB, \ell) \big) \Big), & a = \actB,\\
        0, & \mathrm{otherwise},
    \end{cases}
\end{equation}
and
\begin{equation}
    w_X^V(a, \ell) = 
    \begin{cases}
        -\log_2\Big( \Pr \big( E \mid r_X^{(\ell)} ; \sigma_X^A(\actX, \ell) \big) \Big), & a = \actX, \\
        -\log_2\Big( \Pr \big( E \mid r_X^{(\ell)} ; \sigma_X^A(\actB, \ell) \big) \Big), & a = \actB, \\ 
        0, & \mathrm{otherwise}.
    \end{cases}
\end{equation}
Here $r_Z^{(\ell)}$ and $r_X^{(\ell)}$ denote the measurement outcomes of the ancilla qubits used for the $\ell$-th $Z$-type and $X$-type surface-code stabilizer measurements, respectively.

After constructing the decoding graphs, we apply the MWPM algorithm to each of them. 
For each matched pair, the corresponding edge in the decoding graph represents a shortest path between the two detection events in the syndrome graph. 
We then apply the corresponding GKP logical Pauli corrections to the data qubits associated with the space-like edges along the selected shortest paths. 
The selected paths in the $Z$-type syndrome graph determine the $\bar{X}_{\mathrm{gkp}}$ corrections, while those in the $X$-type syndrome graph determine the $\bar{Z}_{\mathrm{gkp}}$ corrections.

\section{Simulation results}\label{app:result}
In this appendix, we provide additional simulation results for the proposed adaptive skipping scheme.

\subsection{Additional finite-round performance}\label{app:finite}
In Sec.~\ref{subsec:finite}, we fixed the idle-noise variance to $2\sigma_{\mathrm{idle}}^2 = \sigma^2$, so that idle noise is smaller
than circuit noise. 
Here, we present additional finite-round simulation results for two other choices of the idle-noise strength to examine how the relative strength of idle and circuit noise affects the performance of the full-measurement baseline, the all-skip strategy, and the adaptive skipping scheme. 
In both settings, the simulation procedure and decoding method are identical to those in Sec.~\ref{subsec:finite}, except for the idle-noise parameter. 
Since $\sigma = 0$ implies $\sigma_{\mathrm{idle}} = 0$ under all three choices, case~(a) yields identical results regardless of the idle-noise setting. 
We therefore focus on cases~(b) and~(c).

\begin{figure}[tbh]
    \centering
    \includegraphics[width=\linewidth]{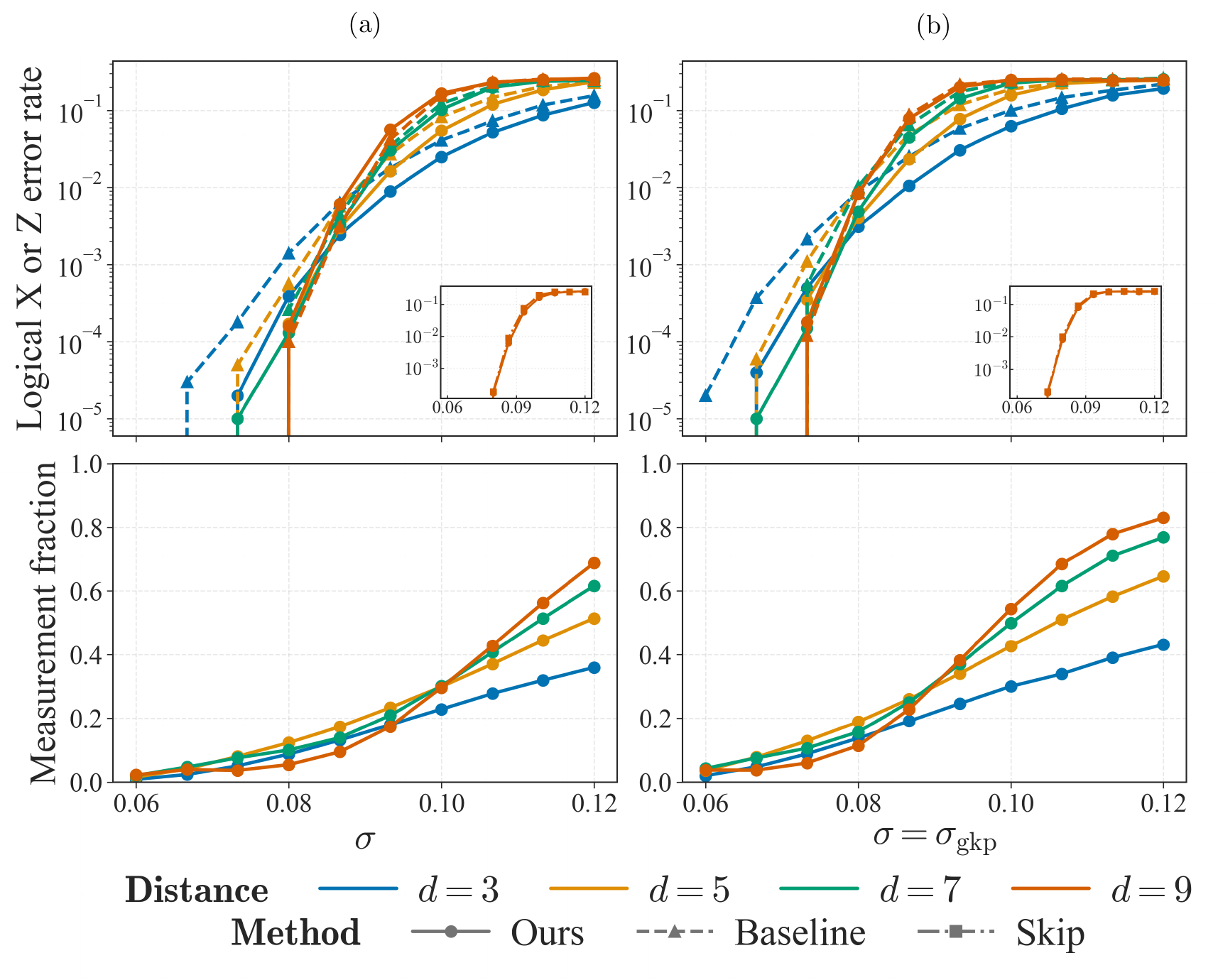}
    \caption{
    Finite-round performance of the adaptive skipping scheme with $\sigma = \sigma_{\mathrm{idle}}$.
    We compare two noise regimes: 
    (a) $\sigma_{\mathrm{gkp}}=0$, and
    (b) $\sigma = \sigma_{\mathrm{gkp}}$.
    Solid lines denote the adaptive skipping scheme, dashed lines denote the full-measurement baseline, and the insets compare the adaptive skipping scheme with the all-skip strategy for $d=9$.
    }
    \label{fig:app-same}
\end{figure}

\begin{figure}[tbh]
    \centering
    \includegraphics[width=\linewidth]{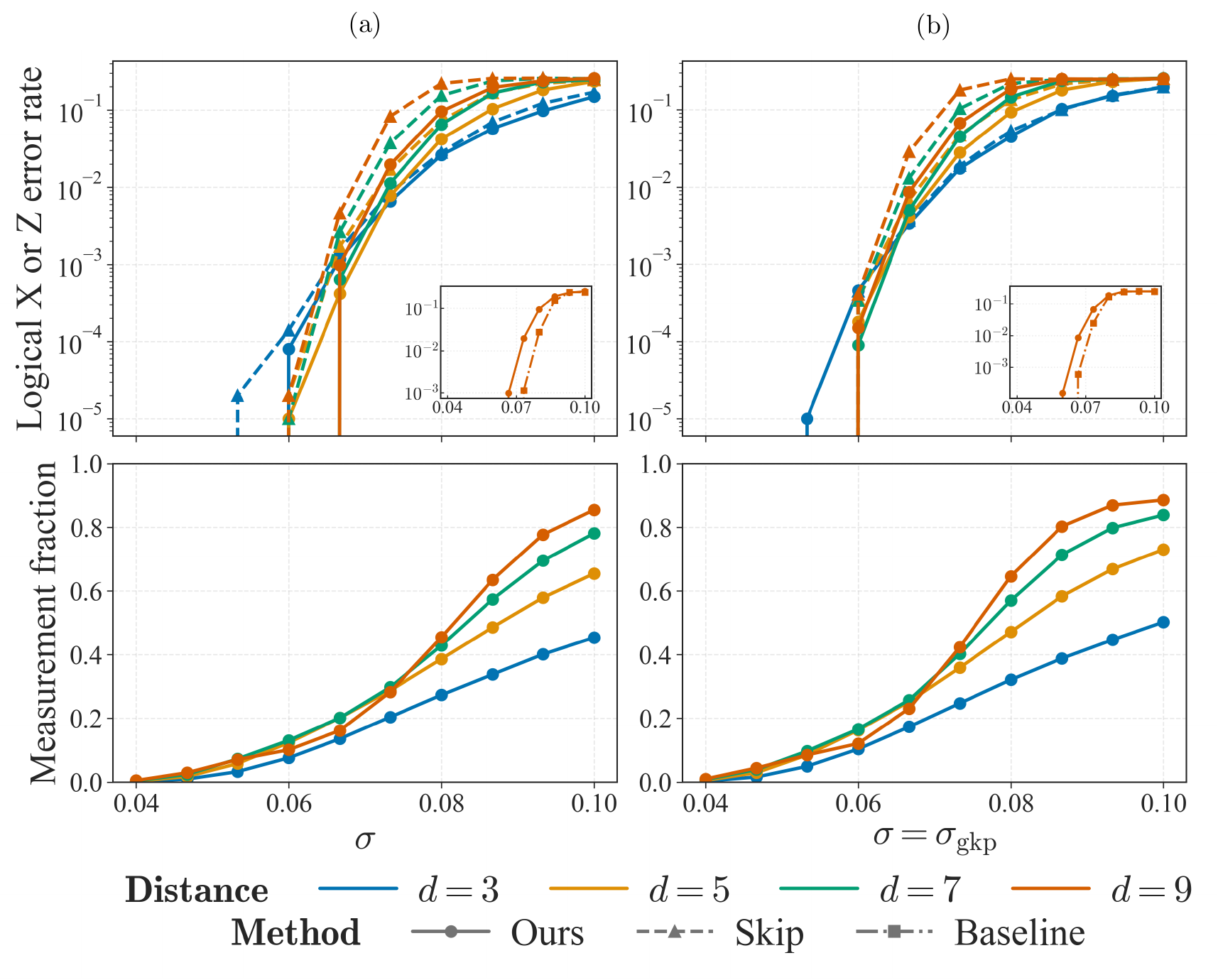}
    \caption{
    Finite-round performance of the adaptive skipping scheme with $2\sigma ^2= \sigma_{\mathrm{idle}}^2$.
    We compare two noise regimes: 
    (a) $\sigma_{\mathrm{gkp}}=0$, and
    (b) $\sigma = \sigma_{\mathrm{gkp}}$.
    Solid lines denote the adaptive skipping scheme, dashed lines denote the all-skip strategy, and the insets compare the adaptive skipping scheme with the full-measurement baseline for $d=9$.
    }
    \label{fig:app-worse}
\end{figure}
Figure~\ref{fig:app-same} shows the results for $\sigma_{\mathrm{idle}} = \sigma$, where idle noise and circuit noise have equal strength. 
Compared with the $\sigma_{\mathrm{idle}} < \sigma$ regime in Sec.~\ref{subsec:finite}, the measurement fraction is higher, indicating that fewer stabilizer measurements are skipped when idle noise is no longer suppressed relative to circuit noise. 
Nevertheless, the adaptive skipping scheme still achieves logical error rates comparable to or lower than those of the full-measurement baseline. 
This indicates that selectively avoiding noisy syndrome extraction can remain beneficial even when idle noise is comparable to circuit noise.

Figure~\ref{fig:app-worse} shows the results for $\sigma_{\mathrm{idle}}^2 = 2\sigma^2$, where idle noise is larger than circuit noise. 
In this regime, the all-skip strategy yields higher logical error rates than the full-measurement baseline, because the idle noise accumulated during skipped intervals exceeds the noise introduced by syndrome extraction. 
By contrast, the adaptive skipping scheme achieves lower logical error rates than the all-skip strategy by increasing the measurement frequency in response to the elevated idle noise. 
As indicated by the measurement fraction, the scheme measures a substantially larger fraction of stabilizers than in the $\sigma_{\mathrm{idle}} \leq \sigma$ settings. 
The insets compare the adaptive skipping scheme with the full-measurement baseline for $d = 9$. 
The adaptive scheme does not fully match the baseline performance because some rounds are still skipped based on the estimated syndrome reliability, leading to additional idle-noise errors.

\begin{figure}[t]
    \centering
    \includegraphics[width=\linewidth]{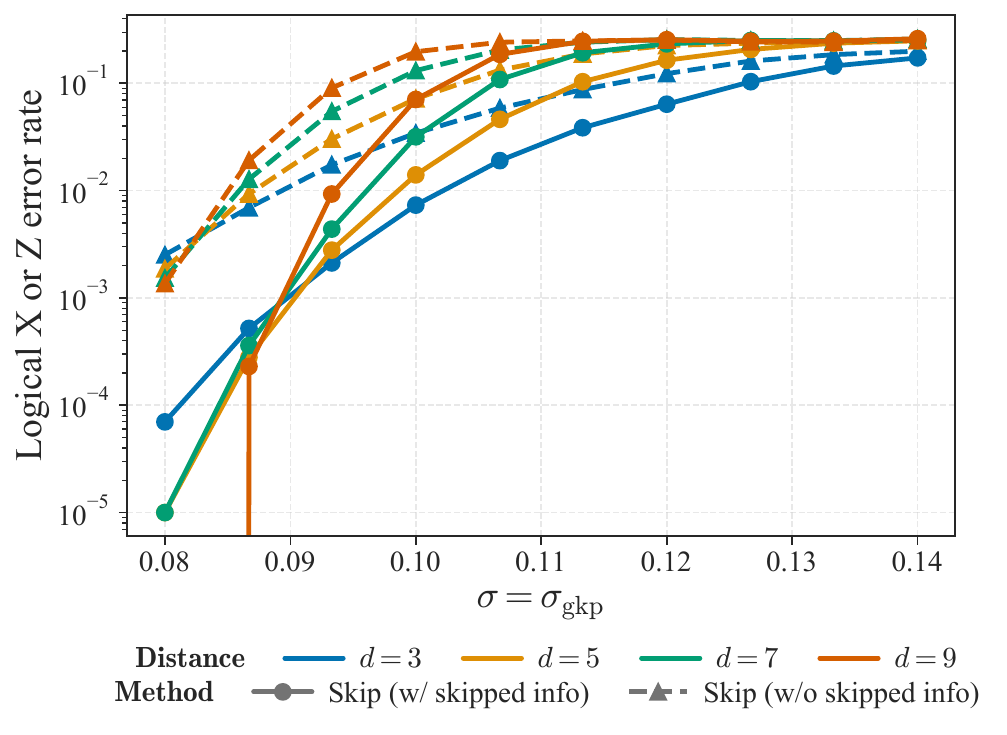}
    \caption{
    Effect of analog information from skipped rounds on the all-skip strategy with $2\h=\c=\g$.
    Solid lines denote decoding with analog information from skipped rounds, and dashed lines denote decoding without analog information from skipped rounds.
    }
    \label{fig:app-skip-analog}
\end{figure}

We also examine the role of analog information from skipped rounds in the decoding performance. 
In the all-skip setting, if the analog information from skipped rounds is discarded, the decoder can only use the two-dimensional syndrome graph from the final noiseless round. 
By contrast, when this information is retained, it can be used to assign space-like edge weights in the three-dimensional syndrome graph, even though no surface-code syndrome is extracted during the skipped rounds. 
Since the time-like edges of skipped rounds carry zero weight, a shortest path can move to a skipped round at no additional time-like cost, traverse a space-like edge in that round, and then return to another time slice. 
Therefore, the space-like edge weights assigned from analog information in skipped rounds can affect the shortest-path distances used to construct the decoding graph. 
Figure~\ref{fig:app-skip-analog} shows this comparison for $2\h = \c = \g$. 
The solid lines correspond to decoding with analog information from skipped rounds, and the dashed lines correspond to decoding without it. 
Notably, discarding the analog information from skipped rounds significantly degrades the performance of the all-skip strategy. 
The insets compare the all-skip strategy without analog information with the full-measurement baseline for $d = 9$. 
The all-skip strategy without analog information yields the highest logical error rates among all methods considered in this setting. 
These results indicate that the analog information obtained during skipped rounds provides valuable guidance to the decoder and plays a critical role in maintaining decoding performance when surface-code syndrome extraction is not performed.

\begin{figure}[t]
    \centering
    \includegraphics[width=\linewidth]{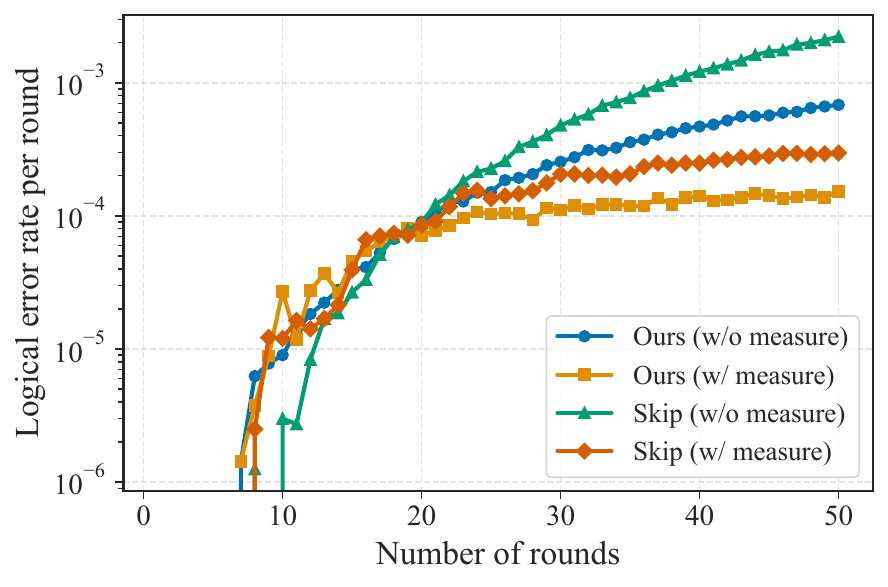}
    \caption{
    Effect of periodic full syndrome extraction on long-time behavior at distance $d=7$.
    The logical error rate per round is shown for the all-skip strategy and the adaptive skipping scheme, with and without forced full surface-code syndrome extraction every $d$ rounds.
    }
    \label{fig:divergence}
\end{figure}

\subsection{Long-time behavior without periodic measurement}\label{app:long}

Next, we consider the role of periodic full-syndrome extraction in long-time simulations. Figure~\ref{fig:divergence} shows the logical error rate per round for the all-skip strategy and the adaptive skipping scheme at distance $d=7$, both with and without periodic full syndrome extraction. Without periodic full syndrome extraction, the logical error rate per round does not converge to a stable value as the number of rounds increases for either strategy. This indicates that periodic full syndrome extraction is necessary to prevent long-time error accumulation.

This behavior is caused by the accumulation of errors during skipped rounds. Although GKP error correction is performed in every round, logical Pauli errors still occur with finite probability, and the total number of such errors grows with the number of rounds. Without periodic full syndrome extraction, the reused syndrome values become increasingly stale, and the MWPM decoder lacks detection events distributed along the time direction to localize errors temporally. As the number of rounds grows, this mismatch becomes more severe, preventing the logical error rate per round from stabilizing. Nevertheless, the adaptive skipping scheme still achieves a lower logical error rate per round than the all-skip strategy. Unlike the all-skip strategy, which never performs surface-code stabilizer measurements during the skipped intervals, the adaptive scheme can perform additional measurements when the reliability of the stored syndrome values becomes sufficiently low. As a result, it can obtain new syndrome information before the reused syndrome history becomes too unreliable.

These results motivate the use of periodic full syndrome extraction adopted in Sec.~\ref{subsec:long}, where forcing a full measurement every $d$ rounds regularly refreshes the surface-code syndrome information and suppresses long-time error accumulation.

\pagebreak
\bibliographystyle{apsrev4-2}
\bibliography{reference.bib}

\end{document}